\documentclass[10pt,aps,prb,twocolumn,amsmath,amssymb,superscriptaddress,nobibnotes]{revtex4-1}
\usepackage{amsmath}
\usepackage{graphicx}
\usepackage{color}
\usepackage{multirow}
\usepackage[caption = false]{subfig}
\usepackage{array}
\usepackage{threeparttable}
\usepackage{mhchem}
\usepackage{mathtools}
\usepackage{siunitx}
\usepackage{centernot}
\usepackage{tikz}
\tikzset{font={\fontsize{11pt}{12}\selectfont}}
\usetikzlibrary{shapes,arrows}
\usepackage{lipsum}
\usepackage{ulem}
\makeatletter

\renewcommand*{\p@subsection}{}

\renewcommand*{\p@subsubsection}{}
\makeatother

\usepackage{hyperref}
\hypersetup{
	linkcolor=blue,          
	citecolor=blue,        
	urlcolor=blue,           
}

\newcommand{\ket}[1]{|#1\rangle}

\newcommand{\inner}[2]{\langle#1|#2\rangle}

\newcommand{\expecth}[3]{\langle#1|#2|#3\rangle}

\begin{document}
\title{Excited states in auxiliary field quantum Monte Carlo}

\author{Ankit Mahajan}
\email{ankitmahajan76@gmail.com}
\affiliation{Department of Chemistry, Columbia University, New York, NY 10027, USA}

\author{Sandeep Sharma}
\email{sanshar@gmail.com}
\affiliation{Department of Chemistry, California Institute of Technology, Pasadena, California 91125, USA}

\author{Shiwei Zhang}
\email{szhang@flatironinstitute.org}
\affiliation{Center for Computational Quantum Physics, Flatiron Institute, New York, New York 10010, USA}

\author{David R. Reichman}
\email{drr2103@columbia.edu}
\affiliation{Department of Chemistry, Columbia University, New York, NY 10027, USA}

\begin{abstract}
We systematically investigate the calculation of excited states in quantum chemistry using auxiliary field quantum Monte Carlo (AFQMC). Symmetry allows targeting of the lowest triplet excited states in AFQMC based on restricted open-shell determinants, effectively as a ground-state calculation. For open-shell singlet states, excited state calculations can be stabilized with the appropriate trial states, but their quality can have a larger effect on the accuracy in AFQMC. We find that active space-based configuration interaction trial states are often not sufficient to obtain accurate results. We instead use truncated equation of motion coupled cluster with single and double excitations (EOM-CCSD) as trial states. We benchmark the performance of these calculations on a set of small and medium molecules and polyacenes, focusing on predominantly single excitations. We find that the AFQMC results, obtained at a per-sample cost scaling of \(O(N^6)\), are systematically more accurate than those obtained using EOM-CCSD, reducing excitation energy errors by approximately half for open-shell singlets. In regimes where EOM-CC triples are impractical, these results position AFQMC as a scalable, higher-accuracy complement for low-lying excited states.

\end{abstract}
\maketitle

\section{Introduction}

An accurate description of optically excited electronic states is essential for understanding diverse chemical and physical processes, from natural phenomena like photosynthesis\cite{cheng2009dynamics} and vision\cite{polli2010conical} to technological applications including photovoltaics and light-emitting diodes.\cite{mathew2014dye} Despite significant advances, computational methods for excited states remain less accurate than those for ground states. This accuracy gap matters: even small errors of a few kcal/mol in excited state potential energy surfaces can radically alter the nonadiabatic dynamics following an excitation. Furthermore, numerical calculations are crucial for studying processes that access dark excited states, which are difficult to probe experimentally. Consequently, developing methods that balance high accuracy with reasonable computational cost remains a central challenge in quantum chemistry.\cite{gonzalez2012progress}

Existing techniques for calculating excited states can be broadly classified along two axes. The first axis distinguishes between wave function and density functional theory (DFT) based methods. Wave function methods offer higher accuracy and can often be systematically improved by increasing the level of theory.\cite{mcweeny1989methods,koch1990coupled,andersson1992second,foresman1992toward,stanton1993equation,angeli2001introduction,gilbert2008self,shea2018communication} However, the steep computational cost of accurate wave function methods limits them to systems with tens of atoms. Many-body Green's function techniques\cite{tiago2006optical,trofimov1995efficient} represent an alternative approach that has recently gained renewed interest. In contrast, density functional theory (DFT) methods\cite{runge1984density,dreuw2005single,hait2021orbital} can handle much larger systems containing hundreds of atoms at reasonable computational cost, but are generally less accurate. Good performance requires careful selection of the exchange-correlation functional, which is often difficult to achieve without input from experiment or benchmark data from more accurate computations.

The second axis differentiates linear response from state-specific methods.  Linear response approaches, such as configuration interaction singles (CIS),\cite{foresman1992toward} time-dependent DFT (TDDFT),\cite{runge1984density,dreuw2005single} and equation-of-motion coupled cluster (EOM-CC),\cite{koch1990coupled,stanton1993equation} calculate multiple excited states simultaneously by building excitations on top of the ground state. While theoretically exact in certain limits, in practice, with approximations and truncations, they tend to be biased toward the ground state. State-specific methods, in contrast, optimize each excited state separately. Examples include state-specific HF,\cite{goddard1969excited,gilbert2008self} DFT,\cite{hait2021orbital} and complete active space based multiconfigurational methods,\cite{andersson1992second,angeli2001introduction} among others.\cite{shea2018communication,lee2019excited,kossoski2023state} State-specific methods typically achieve higher accuracy, particularly when orbital relaxation is crucial, as in charge transfer states. But these methods often require careful selection of initial guesses and optimization strategies, making them less black-box than linear response methods.

Diffusion Monte Carlo (DMC) techniques\cite{foulkes2001quantum} rank among the most accurate methods for ground state calculations, yet their application to excited states has remained limited until recently.\cite{schautz2004excitations,drummond2005electron,zimmerman2009excited,dubecky2010ground,scemama2018excitation,pineda2019excited} Previous studies have revealed that results can depend sensitively on the trial state selection.\cite{schautz2004excitations,scemama2018excitation} Recent algorithmic developments\cite{filippi2016simple} enable the use of large selected configuration interaction (sCI)\cite{Huron1973,Holmes2016b,tubman2016deterministic} expansions as trial states, allowing systematic convergence of fixed-node errors. Alternatively, more scalable approaches employ techniques to exploit cancellation of the fixed-node error between ground and excited state calculations, keeping the trial states compact.\cite{robinson2017excitation,scemama2018excitation,pineda2019excited,dash2019excited}

In this work, we explore auxiliary field quantum Monte Carlo (AFQMC)\cite{zhang2003quantum} as a promising approach to achieve both high accuracy and computational efficiency for excited state calculations. AFQMC has advanced significantly for ground state \textit{ab initio} calculations,\cite{shee2019achieving,motta2018ab,shi2021some,lee2022twenty} particularly through improvements in trial states.\cite{chang2016auxiliary,lee2020utilizing,mahajan2022selected,mahajan2025beyond,danilov2025capturing,xiao2025implementingadvancedtrialwave} However, excited states remain largely unexplored, with a few exceptions.\cite{purwanto2009excited,ma2013excited} Here, we investigate AFQMC performance with multiconfigurational trial states for excited state calculations, focusing on single excitations. As with other methods, symmetry is essential for targeting low-lying excited states in AFQMC. But for states that cannot be targeted based on symmetry alone, appropriately chosen trial wave functions can still enable stable excited-state calculations via the constraint.

We examine CI trial states derived from truncated coupled cluster (CC) wave functions. Building on the recently developed CISD (configuration interaction singles and doubles)
trial state method, which achieved remarkable ground-state accuracy by truncating CCSD wave functions,\cite{mahajan2025beyond} we extend this approach to excited states. We present an efficient algorithm for using truncated EOM-CCSD trial states, termed EOM-CISD, in AFQMC calculations. Despite containing up to quadruple excitations, the per-sample cost of energy calculations with this trial state scales as \(O(N^6)\), compared to \(O(N^5)\) for CISD trials. We also consider a cheaper variant that excludes quadruples, leading to a per-sample cost scaling of \(O(N^5)\). We analyze these approaches in detail and benchmark their performance on small and medium-sized molecules from the QUEST datasets,\cite{loos2020quest} focusing on triplet and open-shell singlet excitations.

In Section \ref{sec:theory}, we outline the theory of AFQMC and the algorithms used to evaluate the trial-dependent quantities. In Section \ref{sec:results}, we present results for a set of small to medium-sized molecules and polyacenes. We conclude with a summary and discussion of future directions in Section \ref{sec:conclusion}.

\section{Theory}\label{sec:theory}
We use the quantum chemistry Hamiltonian given by
\begin{equation}
	H = h_{pq}a_{p\sigma}^{\dagger}a_{q\sigma} + \frac{1}{2}L^{\gamma}_{pq}L^{\gamma}_{rs}a_{p\sigma}^{\dagger}a_{r\lambda}^{\dagger}a_{s\lambda}a_{q\sigma},
\end{equation}
where \(h_{pq}\) are one-electron integrals and \(L^{\gamma}_{pq}\) are Cholesky-decomposed two-electron integrals in an orthonormal orbital basis, and \(\{a_{p\sigma}^{\dagger}\}\) and \(\{a_{p\sigma}\}\) are electronic creation and annihilation operators, respectively, with \(\sigma\) and \(\lambda\) being spin indices. We use the convention of summing over repeated indices throughout this article. For later convenience, we rearrange the terms in the Hamiltonian as

\begin{equation}
   H = H_1 + H_2 = h'_{pq}a_{p\sigma}^{\dagger}a_{q\sigma} + \frac{1}{2}\left(L^{\gamma}_{pq}a_{p\sigma}^{\dagger}a_{q\sigma}\right)\left(L^{\gamma}_{rs}a_{r\lambda}^{\dagger}a_{s\lambda}\right),\label{eq:ham}
\end{equation}
where \(h'_{pq}\) are modified one-electron integrals defined as
\begin{equation}
   h'_{pq} = h_{pq} - \frac{1}{2}L^{\gamma}_{pr}L^{\gamma}_{rq}.
\end{equation}
We also define the one-body operators 
\begin{equation}
   v_{\gamma} = L^{\gamma}_{pq}a_{p\sigma}^{\dagger}a_{q\sigma}.
\end{equation} 
We denote the number of electrons by \(N\), the number of orbitals by \(M\), and the number of Cholesky vectors by \(X\). Below, we present the formulation of AFQMC, emphasizing the aspects relevant to excited state calculations. A more thorough discussion can be found in various review articles.\cite{motta2017computation,shi2021some}

\subsection{Targeting excited states in AFQMC}\label{sec:targeting_excited_states}
In AFQMC, the ground state, \(\ket{\Psi_0}\), is obtained by applying an exponential form of the projector onto an initial state, Slater determinant \(\ket{\phi_0}\), as
\begin{equation}
   e^{-\tau H}\ket{\phi_0} \xrightarrow{\tau\rightarrow\infty}\ket{\Psi_0},
\end{equation}
where \(\tau\) is the imaginary time, and we assume \(\inner{\phi_0}{\Psi_0}\neq 0\). To act the propagator onto a state, we first Trotterize it as
\begin{equation}
   e^{-\tau H} = \prod_{i=1}^{N_t}\left(e^{-\frac{\Delta\tau H_1}{2}}e^{-\Delta\tau H_2}e^{-\frac{\Delta\tau H_1}{2}}\right) + O(\Delta\tau^2),
\end{equation}
where we have divided the imaginary time \(\tau\) into \(N_t\) time slices of size \(\Delta\tau = \tau/N_t\). The two-body operator, which we wrote as a sum of squares in Eq.~\ref{eq:ham}, can be expressed using the Hubbard-Stratonovich transformation as
\begin{equation}
      e^{-\Delta\tau H_2} = \int d\mathbf{x}\ e^{\frac{-\mathbf{x}^2}{2}}e^{i\sqrt{\Delta\tau}x_{\gamma}v_{\gamma}} + O(\Delta\tau^2),\label{eq:hs}
\end{equation}
where \(\mathbf{x}\) is the vector of scalar auxiliary fields \(x_{\gamma}\). Note that the \(O(\Delta\tau^2)\) error is a result of \([v_{\gamma}, v_{\lambda}] \neq 0\), for \(\gamma\neq \lambda\), in general. Error terms linear in \(\Delta\tau\) vanish identically after performing the integral. Thus, we have expressed the two-body projector as an integral over one-body exponentials. Thouless' theorem allows efficient calculation of the action of the one-body exponential operators on a Slater determinant, producing a new Slater determinant as
\begin{equation}
   \ket{\phi'} = e^{-O_{pq}a_p^{\dagger}a_q}\ket{\phi}, 
\end{equation}
with
\begin{equation}
   \phi' = O\phi,
\end{equation}
where \(\phi\) and \(\phi'\) denote the coefficient matrices of the corresponding Slater determinants. Therefore, by sampling the integral over auxiliary fields in Eq.~\ref{eq:hs}, the propagated wave function is expressed as a sum of Slater determinants
\begin{equation}
   e^{-\tau H}\ket{\phi_0} \approx \sum_i \ket{\phi_i}.
\end{equation}
The procedure outlined so far does not introduce any bias in the calculation and is known as free projection (mean field subtraction is always used to reduce the noise in practice, see Ref. \citenum{motta2017computation} for details). 

We now consider the role played by Hamiltonian symmetries in free projection. \(S_z\) symmetry is clearly preserved for each walker as the number of up and down electrons is unchanged throughout the propagation. More generally, it preserves the total spin symmetry if restricted or restricted open-shell walkers are used, as the same operators are applied to the up and down parts of the walkers. This allows the targeting of the lowest states in each spin sector by using the high-spin determinant in the desired multiplet as the initial walker. On the other hand, the propagation does not preserve point group symmetry for each walker. While the one-body operator matrix has a block diagonal structure when using a point group symmetry adapted basis, the Cholesky matrices do not.\cite{motta2019hamiltonian} Therefore, even if the initial determinant belongs to a definite irreducible representation of the point group, the propagated Slater determinant, in general, does not. This symmetry is only preserved on average, and therefore, it is not possible to target the lowest states in an arbitrary irreducible representation of the point group. 

The phaseless approximation is commonly used in AFQMC calculations to eliminate the phase problem, which makes free projection calculations exponentially expensive.\cite{zhang2003quantum} This approximation employs a trial state to control the sign or phase of the wave function. It requires an importance sampling transformation of the propagator, leading to a real-valued weight attached to each walker. The propagated state is represented as
\begin{equation}
   \ket{\psi} = \sum_i w_i \frac{\ket{\phi_i}}{\inner{\psi_T}{\phi_i}},
\end{equation}
where \(w_i\) are the weights, and \(\ket{\psi_T}\) is the trial state. The weights are updated at each propagation step as
\begin{equation}
w_{\phi'} = w_{\phi}\Theta\left(\text{Re}\frac{\inner{\psi_T}{\phi'}}{\inner{\psi_T}{\phi}}\right),
\end{equation} 
where \(\Theta\) is the Heaviside step function. This ensures that walkers with larger overlaps with the trial state are weighted more heavily. It also prevents the phase of the walker overlap from changing by more than \(\pi/2\), thus controlling the sign problem at the expense of a bias. The error in energy due to the phaseless bias is determined by the accuracy of the trial state, and it vanishes in the limit of an exact trial state. In practice, a force bias, given by
\begin{equation}
   \bar{v}_{\gamma} = -i\sqrt{\Delta\tau}\frac{\expecth{\psi_T}{v_{\gamma}}{\phi}}{\inner{\psi_T}{\phi}},
\end{equation}
is used to shift the auxiliary field values, so as to keep the walker overlap with the trial state from vanishing, and to reduce the variance of the weights. The mixed estimator of energy is calculated as the weighted average of local energies, given by
\begin{equation}
   E_L[\phi] = \frac{\expecth{\psi_T}{H}{\phi}}{\inner{\psi_T}{\phi}}.
\end{equation}

The phaseless constraint offers a natural way to target excited states that cannot be accessed using symmetry alone in free projection. Since it imposes a boundary condition on free projection based on the trial state, using a trial that approximates the desired excited state allows AFQMC to sample it. We note that this strategy is also employed in fixed-node DMC calculations to target excited states. However, the two methods work in different manifolds, and the accuracy of the fixed-node and the phaseless approximations, using the same trial wave function, can be rather different. Since the phaseless bias is controlled by the quality of the trial state, we now turn to the crucial choice of trial states for excited state calculations.

\subsection{Trial states}\label{sec:trial_states} 

\begin{figure}[htp]
   \centering
   \includegraphics[width=0.4\textwidth]{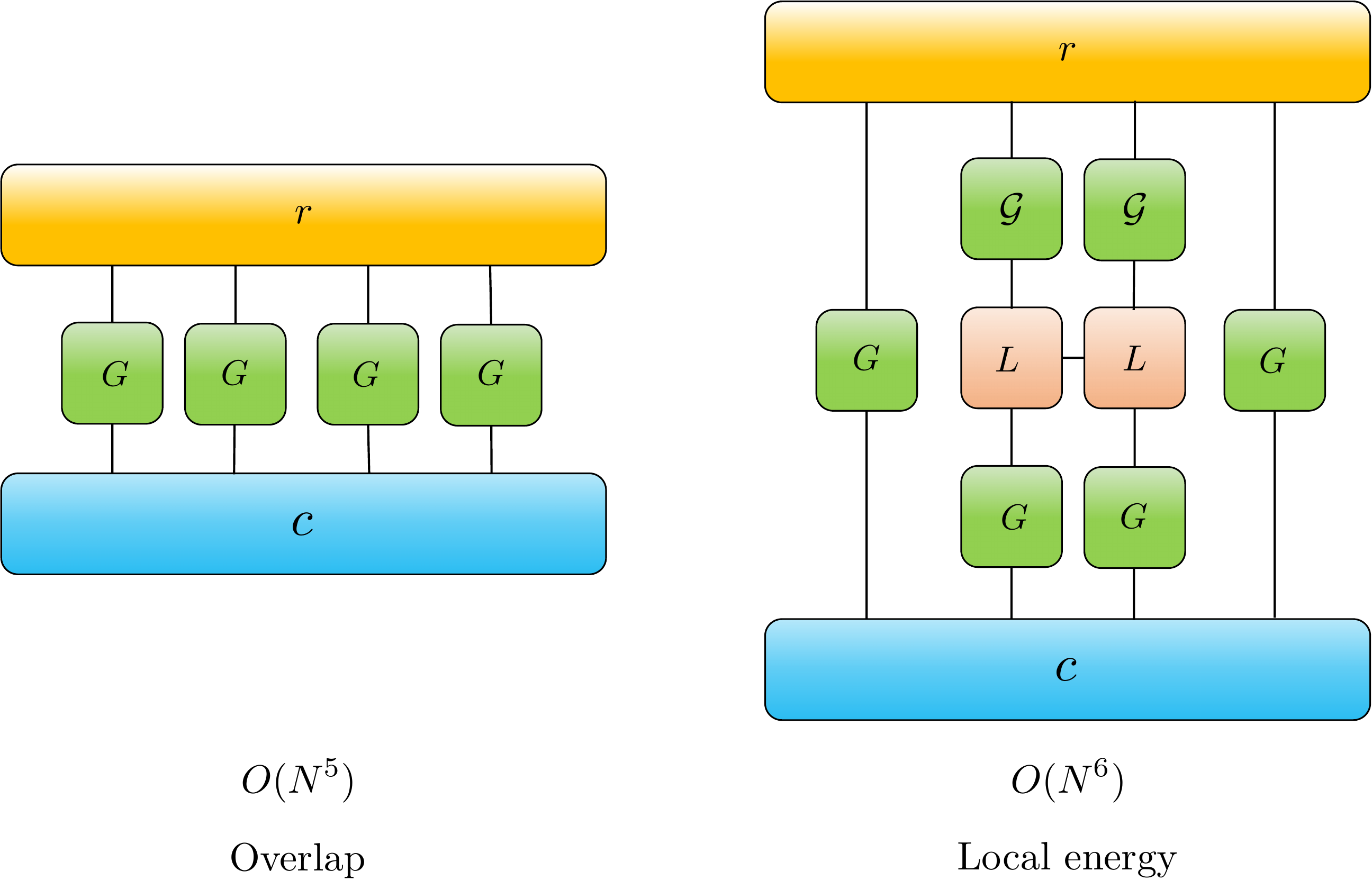}
   \caption{Tensor diagrams for the most expensive terms, involving $R_2C_2$, in the evaluation of overlap (see Eq.~\ref{eq:eom_overlap}) and local energy of the EOM-CISD trial state. Here, \(r\) and \(c\) are EOM and CI amplitudes defined in Eqs.~\ref{eq:eom_amps} and \ref{eq:ci_amps}, \(G\) is the Green's function defined in Eq.~\ref{eq:green}, and the modified Green's function is given by \(\mathcal{G}^p_q = G^p_q - \delta^p_q\).}
   \label{fig:tensors}
\end{figure}
Single determinant trial states, obtained from HF, are commonly used in AFQMC ground state calculations.\cite{motta2018ab} Lowest energy high-spin states of a given multiplicity can be targeted using the restricted open-shell (ROHF) approach. Unrestricted determinants can also be used in some cases, especially in strongly correlated systems, but spin-contamination can be an issue, making the state inappropriate for use as a trial in AFQMC. \(\Delta-\)SCF (self-consistent field) methods\cite{gilbert2008self} allow optimization of orbitals for specific excited states, particularly double excitations and charge transfer states, but we do not consider them here.

Open-shell singlet states do not permit a single determinant wave function, requiring at least two determinants for a correct qualitative description. Configuration interaction states are therefore commonly used to describe such excited states. The simplest such wave function is the CIS state,\cite{foresman1992toward} containing single excitations on top of the ground state HF determinant. Multiconfigurational SCF (MCSCF) methods optimize orbitals along with CI coefficients for configurations inside a small active space. Recent algorithmic developments allow long CI expansions in large active spaces, including those from sCI, to be used as AFQMC trial states.\cite{mahajan2020efficient,mahajan2021taming,mahajan2022selected} 

Besides active space trial states, we have also used CISD wave functions as trial states for ground state calculations in AFQMC.\cite{mahajan2025beyond} These can be used to target the lowest energy states in a given spin sector based on the corresponding ROHF reference determinant. But the ground state closed-shell HF determinant-based CISD is known to perform poorly for open-shell singlet states, as it does not include the triple excitations necessary for orbital relaxation. Instead, we use a truncated EOM-CCSD wave function as a trial state for targeting open-shell singlets.\cite{triplet_eom} This state, EOM-CISD, is given by
\begin{equation}
   \ket{\psi_T} = \left(R_1 + R_2\right) (1 + C_1 + C_2)\ket{\phi_0},
\end{equation}
where \(\ket{\phi_0}\) is the HF determinant, the EOM excitation operators are defined as
\begin{equation}
   R_1 = r^{a}_{i}a_{a\sigma}^{\dagger}a_{i\sigma},\quad R_2 = \frac{1}{2}r^{ab}_{ij}a_{a\sigma}^{\dagger}a_{b\lambda}^{\dagger}a_{j\lambda}a_{i\sigma},\label{eq:eom_amps}
\end{equation}
and the CI operators are defined as
\begin{equation}
   C_1 = c^{a}_{i}a_{a\sigma}^{\dagger}a_{i\sigma},\quad C_2 = \frac{1}{2}c^{ab}_{ij}a_{a\sigma}^{\dagger}a_{b\lambda}^{\dagger}a_{j\lambda}a_{i\sigma}.\label{eq:ci_amps}
\end{equation}
Here, \(i, j\) denote occupied orbitals and \(a, b\) the virtual ones. We only consider states based on spin-restricted orbitals in this work. The \(c\) amplitudes are obtained from the ground state CCSD amplitudes as
\begin{equation}
   c^{a}_{i} = \tau^{a}_{i},\quad c^{ab}_{ij} = \tau^{ab}_{ij} + \tau^{a}_{i}\tau^{b}_{j},
\end{equation}
where \(\tau\) are canonical CCSD amplitudes. The \(r\) amplitudes are obtained from an EOM-CCSD calculation. We also consider an approximation of this trial state where the disconnected quadratic term \(R_2C_2\) is neglected. This reduces the cost scaling of the calculation while retaining the important triple excitations. We refer to this trial state as EOM-CISD\(^{*}\). Excitation energies with the truncated EOM trial states are calculated with respect to AFQMC/CISD ground state energies.

The overlap of the walker with the EOM-CISD trial state is calculated using the generalized Wick's theorem (see Ref. \citenum{mahajan2025beyond} for details) as 

\begin{widetext}
   \begin{equation}\label{eq:eom_overlap}
   \begin{split}
      \frac{\inner{\psi_T}{\phi}}{\inner{\phi_0}{\phi}} &= \frac{\expecth{\phi_0}{\left(r^{a'}_{i'}a_{i'\sigma'}^{\dagger}a_{a'\sigma'}+\frac{1}{2}r^{a'b'}_{i'j'}a_{i'\sigma'}^{\dagger}a_{j'\lambda'}^{\dagger}a_{b'\lambda'}a_{a'\sigma'}\right)\left(1 + c^{a}_{i}a_{i\sigma}^{\dagger}a_{a\sigma} + \frac{1}{2}c^{ab}_{ij}a_{i\sigma}^{\dagger}a_{j\lambda}^{\dagger}a_{b\lambda}a_{a\sigma}\right)}{\phi}}{\inner{\phi_0}{\phi}}\\
      &= O_{R_1} + O_{R_1C_1} + O_{R_1C_2} + O_{R_2} + O_{R_2C_1} + O_{R_2C_2} \\
      O_{R_1} &= 2r^a_i G^i_a \\
      O_{R_1C_1} &= 4r^{a'}_{i'} G^{i'}_{a'} c^a_i G^i_a - 2r^a_i c^s_j G^j_a G^i_s \\
      O_{R_1C_2} &= O_{R_1} \left(2c^{ab}_{ij} G^i_a G^j_b - c^{ab}_{ij} G^i_b G^j_a\right) - 2r^{a'}_{i'} G^i_{a'} \left(2c^{ab}_{ij}G^j_bG^{i'}_a - c^{ab}_{ij}G^j_aG^{i'}_b\right)\\
      O_{R_2} &= 2r^{ab}_{ij} G^i_a G^j_b - r^{ab}_{ij} G^i_b G^j_a \\
      O_{R_2C_1} &= 2c^a_i G^i_aO_{R_2} - 2c^a_i G^{i'}_a\left(2r^{a'b'}_{i'j'} G^{j'}_{b'}G^{i}_{a'} - r^{a'b'}_{i'j'} G^{j'}_{a'}G^{i}_{b'}\right)\\
      O_{R_2C_2} &= O_{R_2} \left(2c^{ab}_{ij} G^i_a G^j_b - c^{ab}_{ij} G^i_b G^j_a\right) - \left(2r^{a'b'}_{i'j'} G^{j'}_{b'}G^{i}_{a'} - r^{a'b'}_{i'j'} G^{j'}_{a'}G^{i}_{b'}\right)\left(2c^{ab}_{ij}G^j_bG^{i'}_a - c^{ab}_{ij}G^j_aG^{i'}_b\right)\\
      & + r^{a'b'}_{i'j'}G^{i}_{a'}G^{j}_{b'}(2c^{ab}_{ij}G^{i'}_{a}G^{j'}_{b} - c^{ab}_{ij}G^{i'}_{b}G^{j'}_{a}),
      \end{split}
   \end{equation}
\end{widetext}
where \(G\) is the Green's function defined as
\begin{equation}
   G^p_q = \frac{\expecth{\phi_0}{a_p^{\dagger}a_q}{\phi}}{\inner{\phi_0}{\phi}}.\label{eq:green}
\end{equation}
Note that the factors of 2 and 4 arise from sums over spin indices. The most expensive term is the one on the last line of \(O_{R_2C_2}\) which scales as \(O(N^3M^2)\). This term is shown in Fig.~\ref{fig:tensors} as a tensor diagram. For the EOM-CISD\(^*\) trial state this term is not evaluated and the cost scaling is given by \(O(N^2M^2)\).

The force bias and the local energy can be evaluated using the algorithmic differentiation (AD) based approach outlined in Refs. \citenum{jiang2024unbiasing} and \citenum{mahajan2025beyond}. Briefly, the force bias is written as 
\begin{equation}\label{eq:force_bias_auto}
   \bar{v}_{\gamma} = -i\sqrt{\Delta\tau}\frac{\partial}{\partial x_{\gamma}}\frac{\expecth{\psi_T}{\exp\left(x_{\gamma'}L^{\gamma'}_{pq}a_p^{\dagger}a_q\right)}{\phi}}{\inner{\psi_T}{\phi}}\bigg\vert_{\mathbf{x}=0}.
\end{equation} 
This can then be calculated as the gradient of the overlap between a rotated walker and the trial state, which can be obtained at the same cost scaling as the overlap using reverse mode automatic differentiation. The total cost of evaluating the force bias for an EOM-CISD trial state is \(O(N^3M^2)\). The one-body part of the local energy can be evaluated similarly to the force bias. We write the two-body part as
\begin{equation}
   E_L = \frac{1}{2}\frac{\partial^2}{\partial x^2}\sum_{\gamma}\frac{\expecth{\psi_T}{\exp\left(xL^{\gamma}_{pq}a_p^{\dagger}a_q\right)}{\phi}}{\inner{\psi_T}{\phi}}\bigg\vert_{x=0}.
\end{equation}
Again, due to Thouless' theorem, the numerator reduces to an overlap between the trial state and a Slater determinant. Because the derivative involves a single variable, we evaluate it using finite differences at the same cost scaling, given by \(O(XN^3M^2)\) for the EOM-CISD trial state. For the EOM-CISD\(^*\) trial state, the force bias cost is \(O(N^2M^2)\), and the local energy cost is \(O(XN^2M^2)\). 

While this algorithm is convenient to implement within an AD framework based only on the implementation of the overlap, the explicit evaluation of the force bias and local energy using Wick's theorem is more efficient with a smaller prefactor. We found our manual implementation to be 3-5 times faster than the derivative-based method. The most expensive diagram in the explicit evaluation of local energy is shown in Fig.~\ref{fig:tensors}.

\section{Results}\label{sec:results}
In this section, we present benchmark results for the accuracy of AFQMC/EOM-CISD, comparing to theoretical best estimates in most cases. We make comparisons with coupled cluster methods as well as TDDFT. All correlated calculations employ the frozen core approximation. We used PySCF\cite{sun2018pyscf} for all traditional quantum chemistry calculations. We performed AFQMC calculations with our AD-AFQMC code available in a public repository\cite{dqmc_code} using NVIDIA A100 and H200 GPUs along with four AMD Milan CPU cores. HCI (heat bath CI) calculations\cite{Holmes2016b,ShaHolUmr} were performed using Dice.

\subsection{Illustrative example}\label{sec:illustrative_example}
\begin{table}[htp]
   \caption{Excitation energies (in eV) for butadiene singlet and triplet states using the 6-31+G(d) basis. State-specific CASSCF calculations were performed with the (4\(o\), 4\(e\)) \(\pi\) orbital active space.}\label{tab:butadiene}
   \centering
   \begin{threeparttable}
   \begin{tabular}{ccccc}
   \hline
   \hline
   Method &~~& \(^1B_u\) &~~& \(^3B_u\) \\
   \hline
   CIS && 6.30 && 2.62 \\
   EOM-CCSD && 6.55 && 3.29 \\
   CASSCF && 7.84 && 3.37 \\
   AFQMC/HF && - && 3.49(2) \\
   AFQMC/CIS && 6.89(2) && - \\
   AFQMC/CASSCF && 7.14(3) && 3.39(3) \\
   AFQMC/CISD && - && 3.40(1) \\
   AFQMC/EOM-CISD && 6.46(1) && - \\
   \hline
   Exact && 6.41(2) && 3.37(2) \\
   \hline
   \end{tabular}
   \end{threeparttable}
\end{table}

We first analyze the performance of various trial states for the lowest \(^1B_u\) and \(^3B_u\) states of butadiene in the 6-31+G(d) basis set. These are predominantly \(\pi\rightarrow\pi^*\) single valence excitations with \(\%T_1\) values over \(90\%\) for both states. We chose them as representatives of the results of our AFQMC calculations on similar open-shell singlet and triplet excitations in many main group molecules. Selected CI results extrapolated to the FCI limit for both these states were reported in Ref. \citenum{loos2020mountaineering}, providing reliable reference values. We used the same geometry as in that work. 

\begin{figure}[htp]
   \centering
   \includegraphics[width=0.48\textwidth]{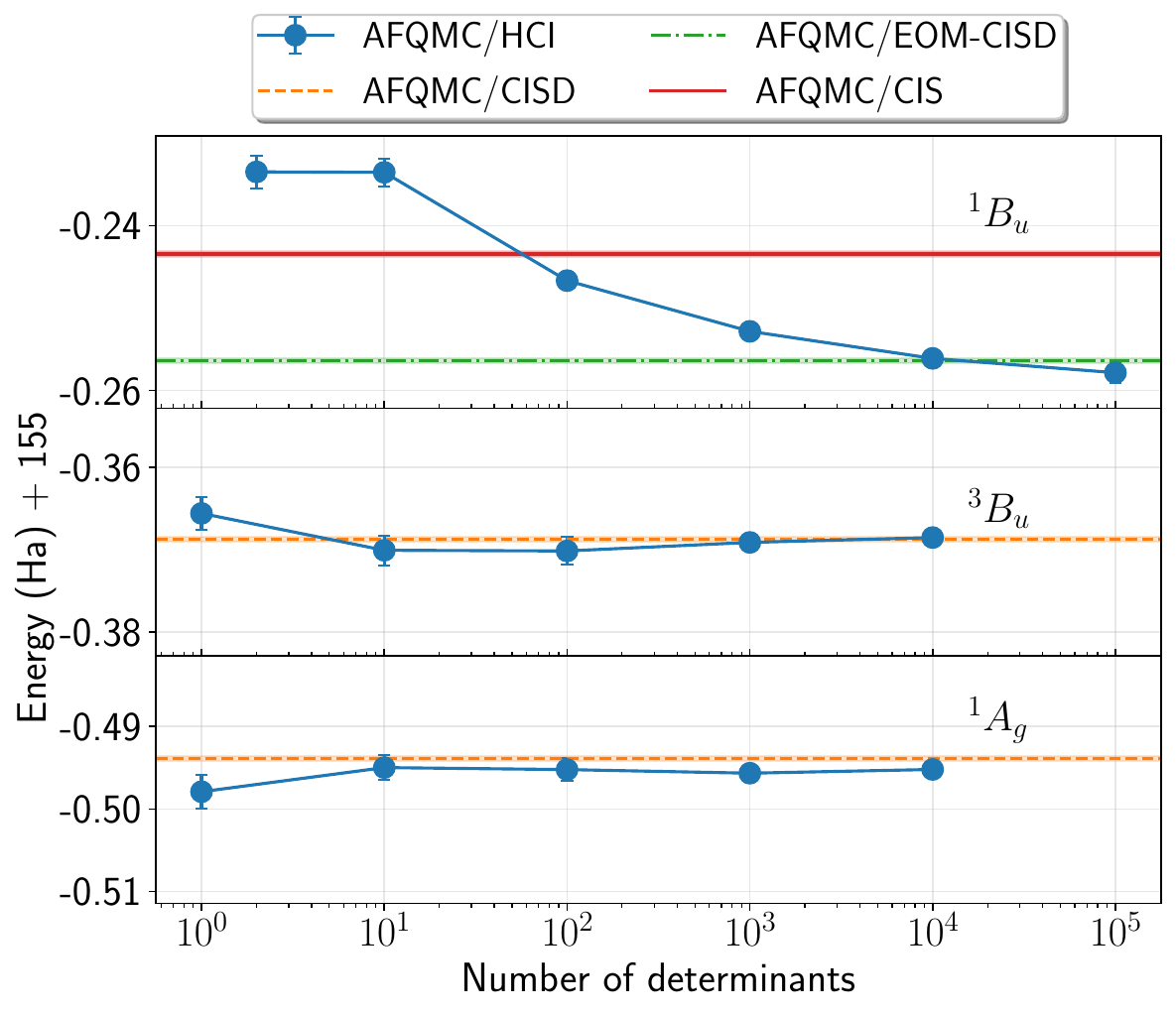}
   \caption{Convergence of absolute AFQMC/HCI energies with respect to the number of determinants in the trial state. We show results for the butadiene ground (\(^1A_g\)), triplet (\(^3B_u\)), and open-shell singlet \(^1B_u\) states. The plot also presents AFQMC energies with the CISD trial for ground and triplet states; CIS, and EOM-CISD trials for the singlet state.}\label{fig:butadiene}
\end{figure}

\subsubsection*{\(^3B_u\) state}\label{sec:butadiene_triplet}
Table \ref{tab:butadiene} shows the excitation energies for these states obtained using various methods. CIS underestimates the excitation energy for the \(^3B_u\) state significantly by 0.75 eV, while EOM-CCSD only slightly underestimates it by 0.08 eV. We used the simplest (4\(o\), 4\(e\)) \(\pi\) active space for CASSCF calculations, which were performed in a state-specific fashion. Despite the lack of dynamical correlation, CASSCF benefits from a fortuitous cancellation of errors and produces a very accurate excitation energy for the triplet. AFQMC, with either CASSCF or CISD trial state, also leads to excitation energies within error bars of the exact result. Both these trials improve upon the AFQMC/HF result, which overestimates the excitation energy by 0.12(3) eV. Fig.~\ref{fig:butadiene} shows the convergence of absolute AFQMC/HCI energies with respect to the number of determinants in the trial state. We used the 40 lowest energy CASSCF orbitals to form an active space for a crude HCI calculation with \(\epsilon_1=10^{-4}\). Both ground and triplet state energies converge relatively quickly, with AFQMC energies converged to within 1 mH with less than 1000 determinants. Therefore, AFQMC with ROHF, CASSCF, and CISD trial states is reasonably accurate for targeting the low-lying triplet state, with CASSCF and CISD trials within chemical accuracy.

\begin{figure}[htp]
   \centering
   \includegraphics[width=0.48\textwidth]{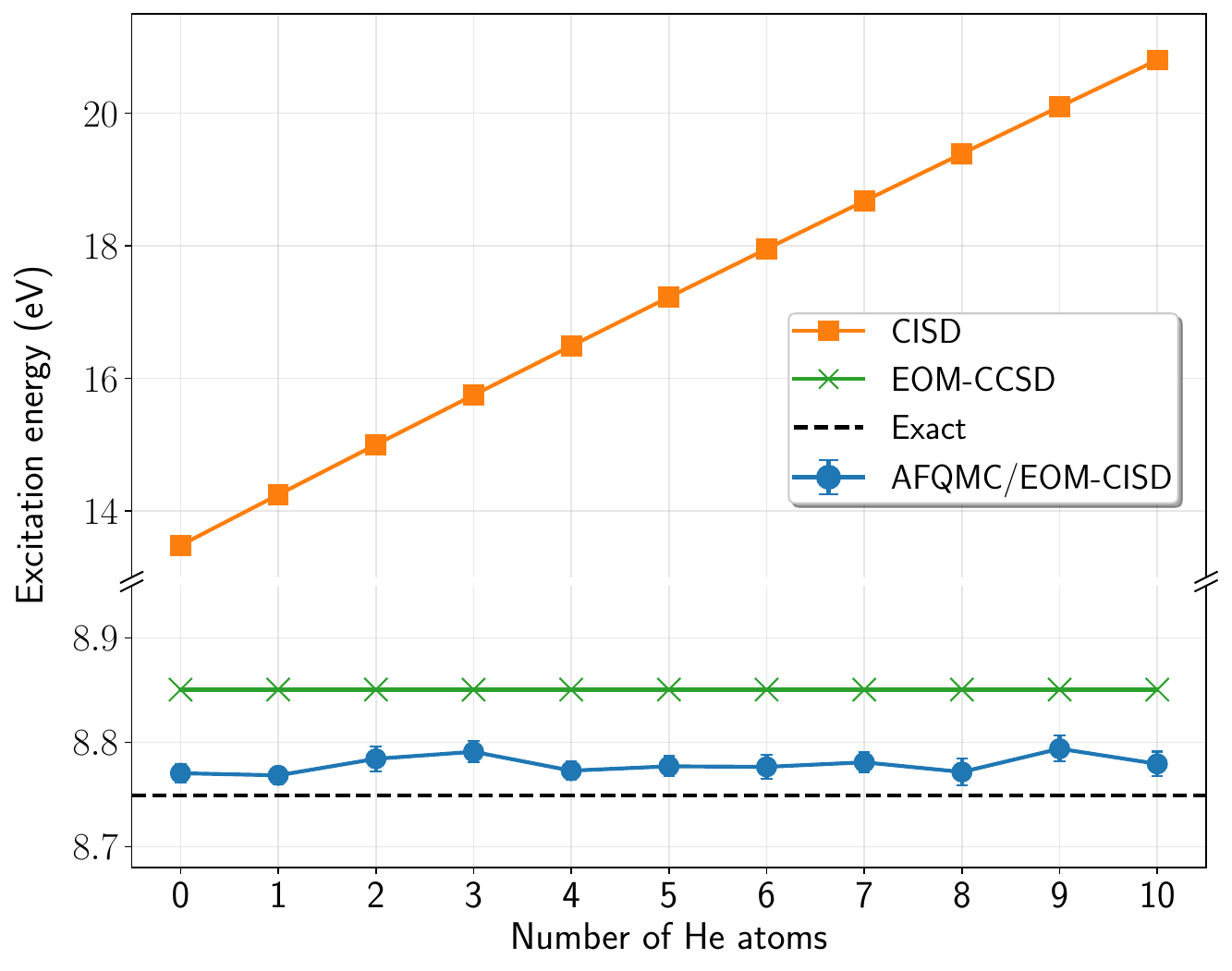}
   \caption{Size intensivity test for the \(^1\Pi\) \(n\rightarrow\pi^*\) excitation of carbon monoxide in the presence of well-separated helium atoms. The excitation energy is shown as a function of the number of helium atoms in the calculation. Note that the \(y\)-axis is broken to accommodate the different energy scales.}\label{fig:size_intensivity}
\end{figure}

\begin{figure*}[htp]
   \centering
   \includegraphics[width=0.95\textwidth]{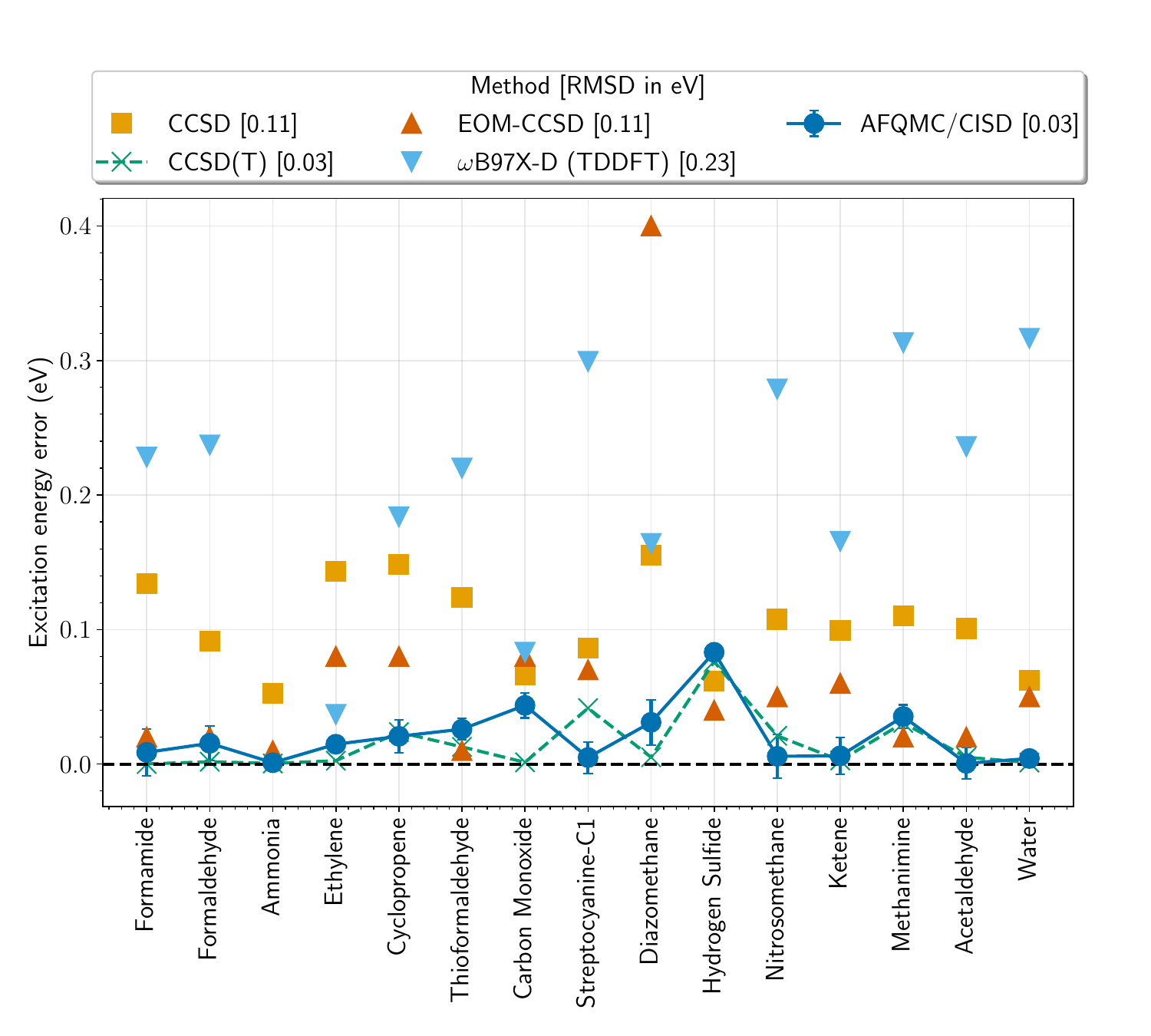}
   \caption{Errors in the excitation energies (in eV) of the lowest triplet excited states of molecules in the QUEST1 dataset\cite{loos2018mountaineering} using the aug-cc-pVTZ basis set. Errors are with respect to the best theoretical estimates reported in Ref. \citenum{loos2018mountaineering}.}\label{fig:triplet_ene}
\end{figure*}

\begin{figure}[htp]
   \centering
   \subfloat{
      \includegraphics[width=0.95\columnwidth]{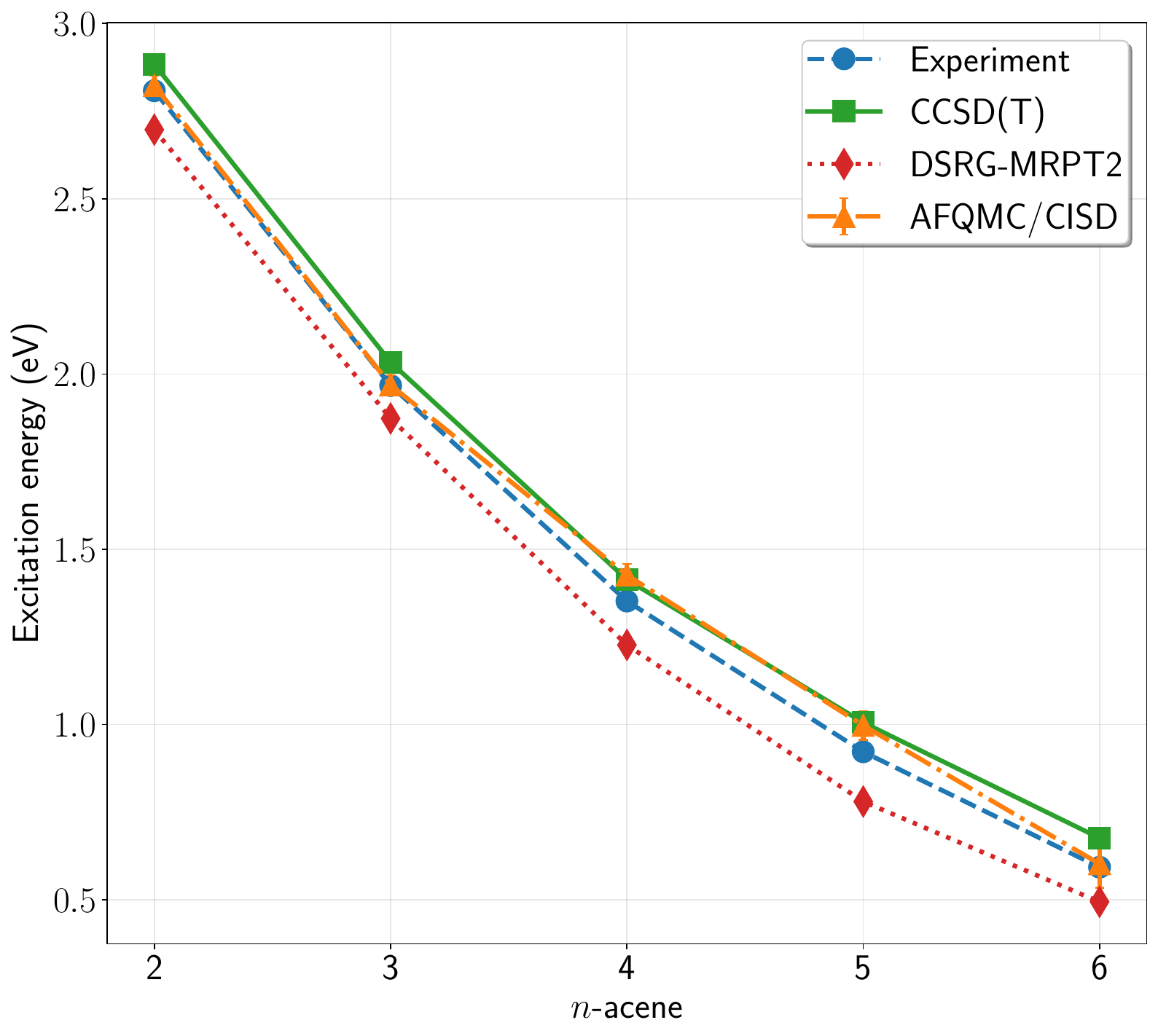}
   }
   \hfill
   \subfloat{
      \small
      \begin{tabular}{lcccccccc}
      \hline
      \hline
      $n$ &~& Experiment &~& AFQMC &~& CCSD(T) &~& DSRG-MRPT2 \\
      \hline
      2 && 2.81 && 2.82(2) && 2.88 && 2.70 \\
      3 && 1.97 && 1.97(2) && 2.03 && 1.87 \\
      4 && 1.35 && 1.43(3) && 1.41 && 1.23 \\
      5 && 0.92 && 1.00(4) && 1.01 && 0.78 \\
      6 && 0.59 && 0.60(6) && 0.68 && 0.49 \\
      \hline
      \end{tabular}%
   }
   \caption{Adiabatic singlet-triplet gaps in polyacenes (in eV) at the complete basis set limit. The table shows numerical values used in the plot. AFQMC values were obtained using CISD trial states. DSRG-MRPT2 values are from Ref.~\citenum{schriber2018combined}. Experimental values have been corrected for zero-point energy.}
   \label{fig:polyacenes}
\end{figure}

\subsubsection*{\(^1B_u\) state}\label{sec:butadiene_singlet}
The \(^1B_u\) open-shell singlet state turns out to be more difficult for almost all the methods. We note that this singlet excitation has been studied extensively with many methods, in part due to the challenge it presents to small \textit{valence} active space methods.\cite{andersson1992second,dallos2004systematic,daday2012full,watson2012excited,dong2019nature} CIS, again somewhat fortuitously, only slightly overestimates the excitation energy by 0.11 eV. EOM-CCSD overestimates it by 0.14 eV, in line with the general trend noted in several previous studies. CASSCF overestimates the excitation energy quite severely by 1.43 eV. This large error has been attributed to the lack of dynamical correlation, leading to significant mixing with Rydberg states. While AFQMC with this trial state cuts this error down by almost half, it is still a significant overestimation by 0.73(5) eV, highlighting the qualitatively poor description of the state by CASSCF with a much too diffuse electronic density due to valence-Rydberg mixing. Convergence of AFQMC/HCI energies for this state, shown in Fig.~\ref{fig:butadiene}, is much slower compared to the ground and triplet states. The energy is nearly converged only with \(10^5\) determinants in the trial state. AFQMC/CIS excitation energy, where we used ground state energy from AFQMC/HF, is also a significant overestimate by 0.48(3) eV. These results highlight the difficulty of converging AFQMC energies with CI expansions for open-shell singlet states, a trend we have observed in several other molecules as well. The large difference in the convergence of AFQMC energies for the ground state and the singlet excited state precludes the use of tricks like variance matching used in DMC excited-state calculations.\cite{robinson2017excitation,pineda2019excited} AFQMC/EOM-CISD performs significantly better, overestimating the excitation by only 0.05(2) eV, a significant improvement over EOM-CCSD. For the singlet excitation, AFQMC/EOM-CISD is the only polynomially scaling approach that delivers chemical accuracy without explicitly including triple excitations.

\subsection{Size intensivity}\label{sec:size_intensivity}
Since AFQMC/EOM-CISD makes use of a truncated CC trial state, it is important to consider possible size extensivity violations in the calculated excitation energies. In Ref. \citenum{mahajan2025beyond}, we found that for ground state calculations with CISD trial states, such violations are relatively small by analyzing the total energy of an ensemble of well-separated molecules. For excitation energies, it is desirable for a method to produce the same excitation energy for a system irrespective of the presence of other noninteracting systems in the calculation. This property is referred to as size intensivity, and EOM-CCSD excitation energies are rigorously size intensive. Similar to Ref. \citenum{tuckman2023excited}, we carry out a numerical test to study intensivity of AFQMC/EOM-CISD excitation energies. We consider the \(^1\Pi\) \(n\rightarrow\pi^*\) valence excitation of carbon monoxide (bond length of 2.142 Bohr) in the cc-pVDZ basis. Fig.~\ref{fig:size_intensivity} shows the excitation energy as a function of the number of well-separated helium atoms in the calculation. The exact energy for a single CO molecule was computed using semistochastic HCI. EOM-CCSD is exactly size-intensive and produces the same excitation energy regardless of the number of helium atoms. CISD excitation energies, on the other hand, are not size-intensive and increase with increasing number of helium atoms. While AFQMC/EOM-CISD is not guaranteed to be size intensive, up to ten helium atoms, we do not see any change in excitation energies within statistical error bars.

\subsection{Triplet states}\label{sec:triplet_states}

In this section, we first present the results for the lowest vertical triplet excitation energies of small molecules in the QUEST1 dataset.\cite{loos2018mountaineering} In AFQMC, we target these states by using CISD trial states based on the high-spin ROHF reference determinant. Fig.~\ref{fig:triplet_ene} shows the errors in the excitation energies of these states calculated in the aug-cc-pVTZ basis. The errors are with respect to the best theoretical estimates reported in Ref. \citenum{loos2018mountaineering}. A recent benchmark\cite{liang2022revisiting} of DFT functionals for TDDFT excitation energies on the QUEST datasets found the hybrid GGA (generalized gradient approximation) \(\omega\)B97X-D functional to be one of the most accurate. We show the TDDFT energies with this functional reported in the benchmark, which show an RMSD error of 0.23 eV. We present both \(\Delta-\)CCSD (based on RHF and ROHF references) and EOM-CCSD excitation energies. Even though they both show an RMSD error of 0.11 eV, EOM-CCSD performs significantly better for all molecules except one outlier, diazomethane. AFQMC/CISD and CCSD(T) both show a much smaller RMSD error of 0.03 eV. Neither method exhibits large outliers, with all errors below 0.1 eV. 

Polyacenes have received a lot of attention in the literature due to the interesting electronic structure of their low-lying states.\cite{bendikov2004oligoacenes,hachmann2007radical,hajgato2011focal,schriber2018combined} In particular, there has been some debate about the nature of the ground state regarding its radical character and spin multiplicity. Experimental results for singlet-triplet (ST) gaps are only available up to hexacene.\cite{siebrand1967radiationless,burgos1977heterofission,angliker1982electronic} We calculated adiabatic ST gaps using AFQMC/CISD in the cc-pVDZ basis using geometries reported in Ref. \citenum{schriber2018combined}. We used the focal point corrections from the coupled cluster study of Ref. \citenum{hajgato2011focal} to obtain values in the basis set limit to compare with experiment. Despite the small basis set, these corrections are all smaller than 0.05 eV for the gaps, suggesting that the error in the correction is relatively small. 

Fig.~\ref{fig:polyacenes} shows the results for the ST gaps of polyacenes up to hexacene with a correlation space of (120\(e\), 418\(o\)). The experimental values shown are corrected for zero-point energy from Ref. \citenum{schriber2018combined}. This paper reported driven similarity renormalization group multireference perturbation theory (DSRG-MRPT2) calculations of the ST gap with large active spaces and basis sets of up to QZ quality. These values follow the experimental trend closely, with a consistent underestimation of the ST gaps by 0.11 to 0.14 eV. CCSD(T) overestimates the gaps with errors that increase slightly with the size of the system. This could be due to the increasing radical character of larger systems. Note that the CCSD(T) values shown here are different from those reported in Ref. \citenum{hajgato2011focal} due to the difference in the geometries used. AFQMC/CISD also slightly overestimates the ST gaps compared to experiment and is in good agreement with CCSD(T). We note that Ref. \citenum{weber2022localized} also reported AFQMC calculations with a CAS trial state and showed similarly good agreement with experiment.

\subsection{Open-shell singlet states}\label{sec:open_shell_singlet_states}
\begin{table*}[htp]
   \caption{Open-shell singlet vertical excitation energies (in eV) of molecules in the QUEST1 dataset using the aug-cc-pVTZ basis set. The last column shows the theoretical best estimate. Bold entries mark the states with the largest absolute deviation from the TBE value for each method. See Fig.~\ref{fig:quest1_hist} for the distribution of errors in excitation energies.}\label{tab:quest}
   \centering
   \begin{threeparttable}
   \begin{tabular}{lcccccccccccc}
   \hline
   \hline
   Molecule &~& Excitation &~& EOM-CCSD &~& DFT (\(\omega\)B97X-D) &~& AFQMC/CISD$^*$ &~& AFQMC/CISD &~& TBE \\
   \hline
   Formamide && \(^1A^{''}\ (V:n\rightarrow\pi^*)\) && 5.69 && 5.64 && 5.75(1) && 5.73(1) && 5.65 \\
   Hydrogen Chloride && \(^1\Pi\) CT && 7.91 && 7.71 && 7.862(4) && 7.843(3) && 7.84 \\
   Formaldehyde && \(^1A_{2}\ (V:n\rightarrow\pi^*)\) && 4.01 && 3.98 && 4.01(1) && 4.00(1) && 3.98 \\
    && \(^1B_{2}\ (R:n\rightarrow3s)\) && 7.23 && 6.95 && 7.27(1) && 7.25(1) && 7.23 \\
    && \(^1B_{2}\ (R:n\rightarrow3p)\) && 8.12 && 7.85 && 8.18(1) && 8.17(1) && 8.13 \\
    && \(^1A_{1}\ (R:n\rightarrow3p)\) && 8.21 && 7.77 && 8.37(1) && 8.32(1) && 8.23 \\
    && \(^1A_{2}\ (R:n\rightarrow3p)\) && 8.65 && 8.27 && 8.80(1) && 8.74(1) && 8.67 \\
    && \(^1A_{1}\ (V:\pi\rightarrow\pi^*)\) && 9.67 && 9.61 && 9.46(1) && 9.46(1) && 9.43 \\
   Ammonia && \(^1A_{2}\ (R:n\rightarrow3s)\) && 6.60 && -- && 6.591(5) && 6.586(4) && 6.59 \\
    && \(^1E\ (R:n\rightarrow3p)\) && 8.15 && -- && 8.164(4) && 8.150(3) && 8.16 \\
    && \(^1A_{1}\ (R:n\rightarrow3p)\) && 9.33 && -- && -- && -- && 9.33 \\
    && \(^1A_{2}\ (R:n\rightarrow3s)\) && 9.95 && -- && 9.928(4) && 9.918(4) && 9.96 \\
   Ethylene && \(^1B_{3u}\ (R:\pi\rightarrow3s)\) && 7.42 && 7.04 && 7.396(6) && 7.384(4) && 7.39 \\
    && \(^1B_{1u}\ (V:\pi\rightarrow\pi^*)\) && 8.02 && 7.95 && 7.876(6) && 7.858(7) && 7.93 \\
    && \(^1B_{1g}\ (R:\pi\rightarrow3p)\) && 8.08 && 7.63 && 8.164(7) && 8.116(6) && 8.08 \\
   Cyclopropene && \(^1B_{1}\ (V:\sigma\rightarrow\pi^*)\) && 6.76 && 6.52 && 6.81(1) && 6.76(1) && 6.68 \\
    && \(^1B_{2}\ (V:\pi\rightarrow\pi^*)\) && 6.86 && 6.61 && 6.72(1) && 6.69(1) && 6.79 \\
   Thioformaldehyde && \(^1A_{2}\ (V:n\rightarrow\pi^*)\) && 2.29 && 2.25 && 2.237(8) && 2.231(7) && 2.22 \\
    && \(^1B_{2}\ (R:n\rightarrow3s)\) && 5.97 && 5.93 && 6.010(8) && 5.953(8) && 5.96 \\
    && \(^1A_{1}\ (V:\pi\rightarrow\pi^*)\) && \textbf{6.63} && 6.70 && 6.48(1) && 6.445(6) && 6.38 \\
   Carbon Monoxide && \(^1\Pi\ (V:n\rightarrow\pi^*)\) && 8.59 && 8.63 && 8.55(1) && 8.533(9) && 8.49 \\
    && \(^1\Sigma^-\ (V:\pi\rightarrow\pi^*)\) && 9.99 && 9.69 && 10.031(8) && 10.002(7) && 9.92 \\
    && \(^1\Delta\ (V:\pi\rightarrow\pi^*)\) && 10.12 && 10.05 && 10.142(9) && 10.125(7) && 10.06 \\
    && \(^1\Sigma^+\ (R)\) && 11.75 && 11.06 && \textbf{11.71(1)} && \textbf{11.71(1)} && 11.52 \\
    && \(^1\Pi\ (R)\) && 11.96 && 11.24 && -- && -- && 11.72 \\
   Dinitrogen && \(^1\Pi_{g}\ (V:n\rightarrow\pi^*)\) && 9.41 && 9.45 && 9.46(1) && 9.43(1) && 9.34 \\
    && \(^1\Sigma_{u}^-\ (V:\pi\rightarrow\pi^*)\) && 10.00 && 9.27 && 9.977(9) && 9.953(9) && 9.88 \\
    && \(^1\Delta_{u}\ (V:\pi\rightarrow\pi^*)\) && 10.44 && 9.75 && 10.42(1) && 10.385(9) && 10.29 \\
    && \(^1\Sigma_{g}^+\ (R)\) && 13.15 && 13.01 && -- && -- && 12.98 \\
    && \(^1\Sigma_{u}^+\ (R)\) && 13.26 && 12.59 && 13.11(1) && 13.05(1) && 13.09 \\
   Acetylene && \(^1\Sigma_{u}^-\ (V:\pi\rightarrow\pi^*)\) && 7.15 && 6.63 && 7.18(1) && 7.17(1) && 7.10 \\
    && \(^1\Delta_{u}\ (V:\pi\rightarrow\pi^*)\) && 7.48 && 6.85 && 7.52(1) && 7.51(1) && 7.44 \\
   Streptocyanine-C1 && \(^1B_{2}\ (V:\pi\rightarrow\pi^*)\) && 7.24 && \textbf{7.96} && 7.17(1) && 7.17(1) && 7.13 \\
   Diazomethane && \(^1A_{2}\ (V:\pi\rightarrow\pi^*)\) && 3.19 && 3.04 && 3.18(1) && 3.14(1) && 3.14 \\
    && \(^1B_{1}\ (R:\pi\rightarrow3s)\) && 5.57 && 5.50 && 5.59(1) && 5.54(1) && 5.54 \\
    && \(^1A_{1}\ (V:\pi\rightarrow\pi^*)\) && 5.94 && 6.07 && 5.87(1) && 5.82(1) && 5.90 \\
   Hydrogen Sulfide && \(^1A_{2}\ (R:n\rightarrow3p)\) && 6.25 && -- && 6.243(3) && 6.224(1) && 6.18 \\
    && \(^1B_{1}\ (R:n\rightarrow3p)\) && 6.29 && -- && 6.261(3) && 6.254(3) && 6.24 \\
   Nitrosomethane && \(^1A^{'}\ (V:n\rightarrow\pi^*)\) && 1.98 && 1.89 && 2.03(1) && 1.99(1) && 1.96 \\
    && \(^1A^{'}\ (R:n\rightarrow3s/3p)\) && 6.43 && -- && -- && -- && 6.29 \\
   Ketene && \(^1A_{2}\ (V:\pi\rightarrow\pi^*)\) && 3.97 && 3.94 && 3.99(1) && 3.93(1) && 3.85 \\
    && \(^1B_{1}\ (R:n\rightarrow3s)\) && 6.09 && 6.00 && 6.10(1) && 6.05(1) && 6.01 \\
    && \(^1A_{1}\ (V:\pi\rightarrow\pi^*)\) && 7.36 && -- && 7.29(1) && 7.28(1) && 7.25 \\
    && \(^1A_{2}\ (R:\pi\rightarrow3p)\) && 7.29 && 6.97 && 7.33(1) && 7.27(1) && 7.18 \\
   Methanimine && \(^1A^{'}\ (V:n\rightarrow\pi^*)\) && 5.28 && 5.22 && 5.249(8) && 5.223(8) && 5.23 \\
   Acetaldehyde && \(^1A^{'}\ (V:n\rightarrow\pi^*)\) && 4.36 && 4.33 && 4.39(1) && 4.36(1) && 4.31 \\
   Water && \(^1B_{1}\ (R:n\rightarrow3s)\) && 7.60 && 7.26 && 7.610(4) && 7.603(4) && 7.62 \\
    && \(^1A_{2}\ (R:n\rightarrow3p)\) && 9.36 && 8.67 && 9.404(4) && 9.383(3) && 9.41 \\
    && \(^1A_{1}\ (R:n\rightarrow3s)\) && 9.96 && 9.45 && 10.006(4) && 9.996(3) && 9.99 \\
   \hline
   Maximum error && && 0.25 && 0.83 && 0.19 && 0.19 \\
   RMSD && && 0.10 && 0.34 && 0.08 && 0.06 \\
   \hline
   \end{tabular}
   \end{threeparttable}
\end{table*}

\begin{figure*}[htp]
   \centering
   \includegraphics[width=0.8\textwidth]{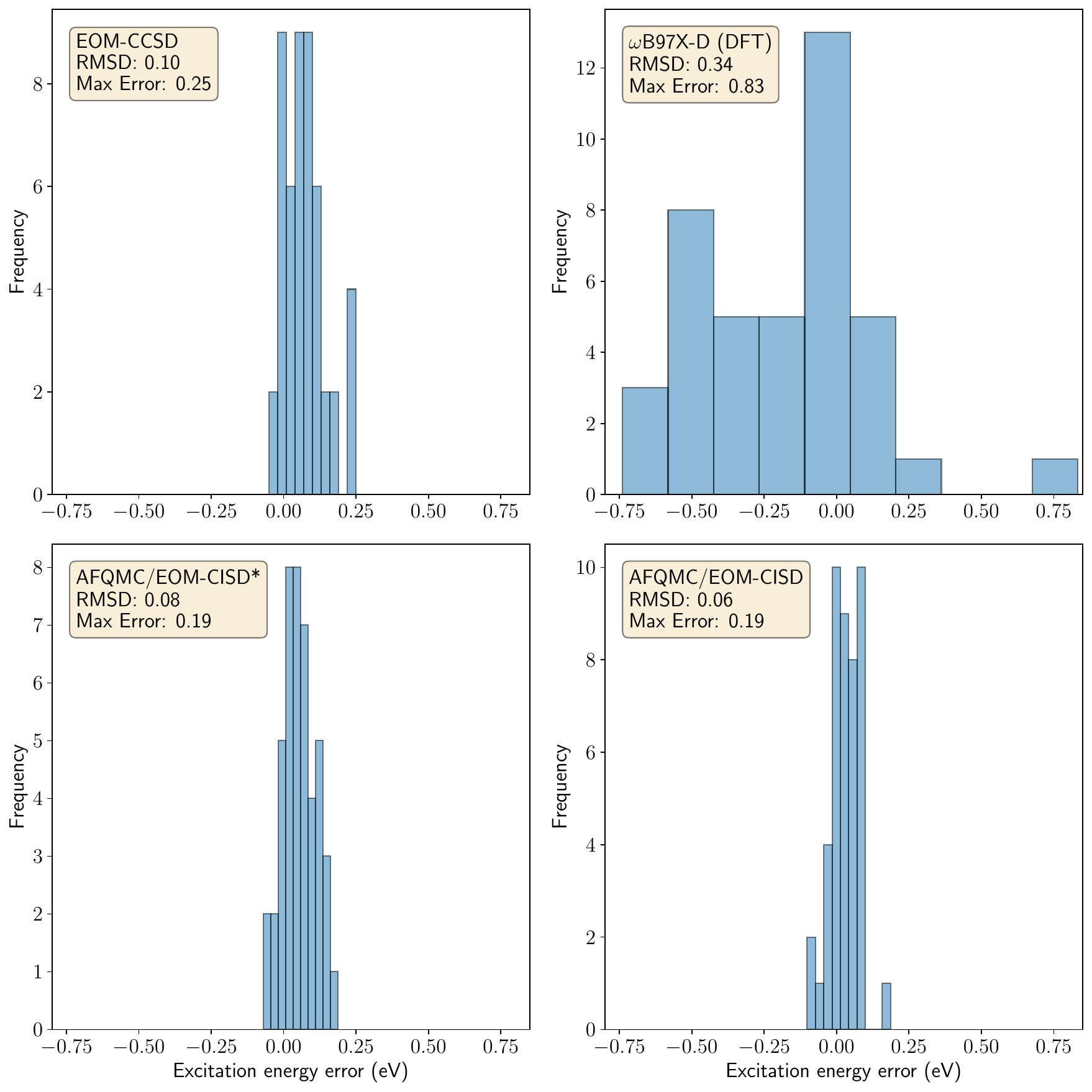}
   \caption{Histogram of errors (in eV) in the open-shell singlet excitation energies for molecules in the QUEST1 dataset shown in Table \ref{tab:quest}. We also show RMS and maximum error for all methods.}\label{fig:quest1_hist}
\end{figure*}
Open-shell singlet excitations present a significant challenge for AFQMC as we discussed in Section \ref{sec:illustrative_example}. In this section, we analyze the performance of AFQMC with EOM-CISD trials for targeting these states in small and medium molecules. 

We calculated the singlet excitation energies of the small molecules in the QUEST1 dataset\cite{loos2018mountaineering} using the aug-cc-pVTZ basis set. We considered 50 vertical open-shell singlet excitations of 18 molecules from this dataset. We excluded states that could not be reliably targeted due to severe state-mixing, for example, higher excitations of formamide. Table \ref{tab:quest} shows the excitation energies calculated using four different methods, while Fig.~\ref{fig:quest1_hist} shows the distribution of errors in the estimates. The errors are overall larger for singlets than for triplets shown in Section \ref{sec:triplet_states}. TDDFT excitation energies using the \(\omega\)B97X-D functional, taken from Ref. \citenum{liang2022revisiting}, show an RMS error of 0.32 eV, with most errors being underestimates. It exhibits an outlier overestimate of 0.83 eV for the \(\pi\rightarrow\pi^*\) excitation to the \(^1B_2\) state of the streptocyanine molecule. Its description of valence states is more accurate (RMSD:0.31) than that of the Rydberg states (RMSD:0.38 eV). 

Of the more expensive wave function methods, EOM-CCSD consistently overestimates excitation energies and has a smaller RMS error of 0.10 eV. AFQMC/EOM-CISD also overestimates excitation energies but cuts the EOM-CCSD RMS error almost in half to 0.06 eV. We also note that the performance of AFQMC/EOM-CISD deteriorates for higher excited states in a given point group irrep. See energies for the series of excited states of formaldehyde in Table \ref{tab:quest}, for example. In the case of four such states, like the excited state of ammonia in the ground state symmetry sector \(^1A_1\), we could not target them reliably with AFQMC. In such cases it may be beneficial to enforce stricter orthogonality to lower lying states in some way, an issue we will analyze in future studies. The cheaper AFQMC/CISD\(^*\), which excludes the disconnected quadruple excitations, exhibits accuracy between EOM-CCSD and AFQMC/EOM-CISD, with an RMS error of 0.08 eV. The three wave function methods show similar accuracy for valence and Rydberg states in this set. Given the difficulty of targeting open-shell singlet states, these results are quite encouraging.

\begin{table*}[htp]
   \caption{Excitation energies (in eV) for some medium-sized molecules in the QUEST3 dataset using the aug-cc-pVTZ basis set. Last column shows the theoretical best estimate reported in Ref. \citenum{loos2020mountaineering}.}\label{tab:quest3}
   \centering
   \begin{threeparttable}
   \begin{tabular}{lcccccccc}
   \hline
   \hline
   Molecule &~~& Excitation &~~& EOM-CCSD &~~& AFQMC/EOM-CISD &~~& TBE \\
   \hline
   Methylenecyclopropene && \(^1B_2\ (V:\pi\rightarrow\pi^*)\) && 4.58 && 4.33(1) && 4.28 \\
   Butadiene && \(^1B_u\ (V:\pi\rightarrow\pi^*)\) && 6.35 && 6.26(2) && 6.22 \\
   Cyclopropenone && \(^1B_1\ (V:n\rightarrow\pi^*)\) && 4.53 && 4.42(2) && 4.26 \\
   Glyoxal && \(^1A_u\ (V:n\rightarrow\pi^*)\) && 3.01 && 3.01(2) && 2.88 \\
   Propynal && \(^1A^{''}\ (V:n\rightarrow\pi^*)\) && 3.94 && 3.88(2) && 3.80 \\
   Cyanoformaldehyde && \(^1A^{''}\ (V:n\rightarrow\pi^*)\) && 3.94 && 3.88(2) && 3.81 \\
   Thiophene && \(^1A_1\ (V:\pi\rightarrow\pi^*)\) && 5.78 && 5.83(3) && 5.64 \\
   && \(^1B_2\ (V:\pi\rightarrow\pi^*)\) && 6.12 && 5.98(3) && 5.98 \\
   Tetrazine && \(^1B_{3u}\ (V:n\rightarrow\pi^*)\) && 2.64 && 2.59(3) && 2.47 \\
   \hline
   \end{tabular}
   \end{threeparttable}
\end{table*}

We also calculated open-shell singlet excitation energies of eight medium sized molecules in the QUEST3 dataset\cite{loos2020mountaineering} using the aug-cc-pVTZ basis set. The results are shown in Table \ref{tab:quest3}. These results follow the same trend as the small molecule case, with both EOM-CCSD and AFQMC/EOM-CISD overestimating the excitation energies. AFQMC/EOM-CISD energies are at least as accurate as EOM-CCSD in all cases except for the \(^1A_1\) \(\pi\rightarrow\pi^*\) excited state of thiophene, where the AFQMC/EOM-CISD error is worse. This is the second singlet state in the \(A_1\) irrep, the lowest state being the ground state. Therefore, this larger error conforms with the ``larger errors for higher states" pattern observed in the QUEST1 dataset. In most other cases, AFQMC/EOM-CISD is significantly more accurate, and its accuracy does not degrade relative to EOM-CCSD for these larger systems. To illustrate the computational requirements, for tetrazine with a \((30e, 316o)\) correlation space, the AFQMC/CISD ground state calculation took about 30 minutes on a single NVIDIA H200 GPU, whereas the AFQMC/EOM-CISD excited state calculation took about 6 hours on the same machine. For reference, the AFQMC/EOM-CISD\(^*\) calculation took about 2.1 hours for the same number of samples.

\section{Conclusion}\label{sec:conclusion}
In this work, we explored the calculation of excited-state energies using AFQMC. For triplet excited states of closed-shell molecules, symmetry enables straightforward targeting using CISD trial states. These calculations achieve accuracy comparable to ground-state AFQMC/CISD results. Our calculations on the singlet-triplet gap of polyacenes showed good agreement with experimental values up to hexacene.

Open-shell singlet states present a greater challenge. When using multideterminantal trial states, AFQMC energies converge slowly as the trial state size increases. To address this limitation, we proposed an EOM-CISD trial state by truncating the EOM-CCSD wave function. This approach yields substantial improvements over EOM-CCSD, reducing errors by approximately half. Despite using a truncated coupled cluster trial state, our tests reveal negligible size-intensivity violations. These results establish AFQMC/EOM-CISD as a scalable approach to systematically improve upon EOM-CCSD without explicitly including triple excitations.

Looking ahead, we plan to extend this methodology to charge transfer and doubly excited states. Given that EOM-CCSD typically produces large overestimates for these excitations,\cite{loos2021reference} state-specific CISD\cite{kossoski2023state} trial states, which effectively capture orbital relaxation effects, may prove more accurate. Additionally, we will investigate orthogonalization strategies\cite{ma2013excited} and alternative phaseless constraints to improve excited-state calculations in AFQMC for more challenging cases. 

\section*{Data availability}
The data supporting the findings of this study are available within the article and at \url{https://github.com/ankit76/afqmc_excited_states_data}. 

\section*{Acknowledgements}
A.M. and D.R.R. were partially supported by NSF CHE-2245592. SS was partially supported by the DOE grant DE-SC0025943. This work used the Delta system at the National Center for Supercomputing Applications through allocation CHE230028 from the Advanced Cyberinfrastructure Coordination Ecosystem: Services and Support (ACCESS) program, which is supported by National Science Foundation grants \#2138259, \#2138286, \#2138307, \#2137603, and \#2138296. A.M. acknowledges the hospitality of the Center for Computational Quantum
Physics at the Flatiron Institute. The Flatiron Institute is a division of the Simons Foundation.



\begin{thebibliography}{73}%
\makeatletter
\providecommand \@ifxundefined [1]{%
 \@ifx{#1\undefined}
}%
\providecommand \@ifnum [1]{%
 \ifnum #1\expandafter \@firstoftwo
 \else \expandafter \@secondoftwo
 \fi
}%
\providecommand \@ifx [1]{%
 \ifx #1\expandafter \@firstoftwo
 \else \expandafter \@secondoftwo
 \fi
}%
\providecommand \natexlab [1]{#1}%
\providecommand \enquote  [1]{``#1''}%
\providecommand \bibnamefont  [1]{#1}%
\providecommand \bibfnamefont [1]{#1}%
\providecommand \citenamefont [1]{#1}%
\providecommand \href@noop [0]{\@secondoftwo}%
\providecommand \href [0]{\begingroup \@sanitize@url \@href}%
\providecommand \@href[1]{\@@startlink{#1}\@@href}%
\providecommand \@@href[1]{\endgroup#1\@@endlink}%
\providecommand \@sanitize@url [0]{\catcode `\\12\catcode `\$12\catcode
  `\&12\catcode `\#12\catcode `\^12\catcode `\_12\catcode `\%12\relax}%
\providecommand \@@startlink[1]{}%
\providecommand \@@endlink[0]{}%
\providecommand \url  [0]{\begingroup\@sanitize@url \@url }%
\providecommand \@url [1]{\endgroup\@href {#1}{\urlprefix }}%
\providecommand \urlprefix  [0]{URL }%
\providecommand \Eprint [0]{\href }%
\providecommand \doibase [0]{http://dx.doi.org/}%
\providecommand \selectlanguage [0]{\@gobble}%
\providecommand \bibinfo  [0]{\@secondoftwo}%
\providecommand \bibfield  [0]{\@secondoftwo}%
\providecommand \translation [1]{[#1]}%
\providecommand \BibitemOpen [0]{}%
\providecommand \bibitemStop [0]{}%
\providecommand \bibitemNoStop [0]{.\EOS\space}%
\providecommand \EOS [0]{\spacefactor3000\relax}%
\providecommand \BibitemShut  [1]{\csname bibitem#1\endcsname}%
\let\auto@bib@innerbib\@empty
\bibitem [{\citenamefont {Cheng}\ and\ \citenamefont
  {Fleming}(2009)}]{cheng2009dynamics}%
  \BibitemOpen
  \bibfield  {author} {\bibinfo {author} {\bibfnamefont {Y.-C.}\ \bibnamefont
  {Cheng}}\ and\ \bibinfo {author} {\bibfnamefont {G.~R.}\ \bibnamefont
  {Fleming}},\ }\href@noop {} {\bibfield  {journal} {\bibinfo  {journal}
  {Annual review of physical chemistry}\ }\textbf {\bibinfo {volume} {60}},\
  \bibinfo {pages} {241} (\bibinfo {year} {2009})}\BibitemShut {NoStop}%
\bibitem [{\citenamefont {Polli}\ \emph {et~al.}(2010)\citenamefont {Polli},
  \citenamefont {Altoe}, \citenamefont {Weingart}, \citenamefont {Spillane},
  \citenamefont {Manzoni}, \citenamefont {Brida}, \citenamefont {Tomasello},
  \citenamefont {Orlandi}, \citenamefont {Kukura}, \citenamefont {Mathies}
  \emph {et~al.}}]{polli2010conical}%
  \BibitemOpen
  \bibfield  {author} {\bibinfo {author} {\bibfnamefont {D.}~\bibnamefont
  {Polli}}, \bibinfo {author} {\bibfnamefont {P.}~\bibnamefont {Altoe}},
  \bibinfo {author} {\bibfnamefont {O.}~\bibnamefont {Weingart}}, \bibinfo
  {author} {\bibfnamefont {K.~M.}\ \bibnamefont {Spillane}}, \bibinfo {author}
  {\bibfnamefont {C.}~\bibnamefont {Manzoni}}, \bibinfo {author} {\bibfnamefont
  {D.}~\bibnamefont {Brida}}, \bibinfo {author} {\bibfnamefont
  {G.}~\bibnamefont {Tomasello}}, \bibinfo {author} {\bibfnamefont
  {G.}~\bibnamefont {Orlandi}}, \bibinfo {author} {\bibfnamefont
  {P.}~\bibnamefont {Kukura}}, \bibinfo {author} {\bibfnamefont {R.~A.}\
  \bibnamefont {Mathies}},  \emph {et~al.},\ }\href@noop {} {\bibfield
  {journal} {\bibinfo  {journal} {Nature}\ }\textbf {\bibinfo {volume} {467}},\
  \bibinfo {pages} {440} (\bibinfo {year} {2010})}\BibitemShut {NoStop}%
\bibitem [{\citenamefont {Mathew}\ \emph {et~al.}(2014)\citenamefont {Mathew},
  \citenamefont {Yella}, \citenamefont {Gao}, \citenamefont {Humphry-Baker},
  \citenamefont {Curchod}, \citenamefont {Ashari-Astani}, \citenamefont
  {Tavernelli}, \citenamefont {Rothlisberger}, \citenamefont {Nazeeruddin},\
  and\ \citenamefont {Gratzel}}]{mathew2014dye}%
  \BibitemOpen
  \bibfield  {author} {\bibinfo {author} {\bibfnamefont {S.}~\bibnamefont
  {Mathew}}, \bibinfo {author} {\bibfnamefont {A.}~\bibnamefont {Yella}},
  \bibinfo {author} {\bibfnamefont {P.}~\bibnamefont {Gao}}, \bibinfo {author}
  {\bibfnamefont {R.}~\bibnamefont {Humphry-Baker}}, \bibinfo {author}
  {\bibfnamefont {B.~F.}\ \bibnamefont {Curchod}}, \bibinfo {author}
  {\bibfnamefont {N.}~\bibnamefont {Ashari-Astani}}, \bibinfo {author}
  {\bibfnamefont {I.}~\bibnamefont {Tavernelli}}, \bibinfo {author}
  {\bibfnamefont {U.}~\bibnamefont {Rothlisberger}}, \bibinfo {author}
  {\bibfnamefont {M.~K.}\ \bibnamefont {Nazeeruddin}}, \ and\ \bibinfo {author}
  {\bibfnamefont {M.}~\bibnamefont {Gratzel}},\ }\href@noop {} {\bibfield
  {journal} {\bibinfo  {journal} {Nature chemistry}\ }\textbf {\bibinfo
  {volume} {6}},\ \bibinfo {pages} {242} (\bibinfo {year} {2014})}\BibitemShut
  {NoStop}%
\bibitem [{\citenamefont {Gonzalez}\ \emph {et~al.}(2012)\citenamefont
  {Gonzalez}, \citenamefont {Escudero},\ and\ \citenamefont
  {Serrano-Andr{\'e}s}}]{gonzalez2012progress}%
  \BibitemOpen
  \bibfield  {author} {\bibinfo {author} {\bibfnamefont {L.}~\bibnamefont
  {Gonzalez}}, \bibinfo {author} {\bibfnamefont {D.}~\bibnamefont {Escudero}},
  \ and\ \bibinfo {author} {\bibfnamefont {L.}~\bibnamefont
  {Serrano-Andr{\'e}s}},\ }\href@noop {} {\bibfield  {journal} {\bibinfo
  {journal} {ChemPhysChem}\ }\textbf {\bibinfo {volume} {13}},\ \bibinfo
  {pages} {28} (\bibinfo {year} {2012})}\BibitemShut {NoStop}%
\bibitem [{\citenamefont {McWeeny}(1989)}]{mcweeny1989methods}%
  \BibitemOpen
  \bibfield  {author} {\bibinfo {author} {\bibfnamefont {R.}~\bibnamefont
  {McWeeny}},\ }\href {https://books.google.com/books?id=qNeeWm6VoCsC} {\emph
  {\bibinfo {title} {Methods of Molecular Quantum Mechanics}}},\ Chemistry,
  Physical and theoretical\ (\bibinfo  {publisher} {Academic Press},\ \bibinfo
  {year} {1989})\BibitemShut {NoStop}%
\bibitem [{\citenamefont {Koch}\ and\ \citenamefont
  {Jorgensen}(1990)}]{koch1990coupled}%
  \BibitemOpen
  \bibfield  {author} {\bibinfo {author} {\bibfnamefont {H.}~\bibnamefont
  {Koch}}\ and\ \bibinfo {author} {\bibfnamefont {P.}~\bibnamefont
  {Jorgensen}},\ }\href@noop {} {\bibfield  {journal} {\bibinfo  {journal}
  {Journal of Chemical Physics}\ }\textbf {\bibinfo {volume} {93}},\ \bibinfo
  {pages} {3333} (\bibinfo {year} {1990})}\BibitemShut {NoStop}%
\bibitem [{\citenamefont {Andersson}\ \emph {et~al.}(1992)\citenamefont
  {Andersson}, \citenamefont {Malmqvist},\ and\ \citenamefont
  {Roos}}]{andersson1992second}%
  \BibitemOpen
  \bibfield  {author} {\bibinfo {author} {\bibfnamefont {K.}~\bibnamefont
  {Andersson}}, \bibinfo {author} {\bibfnamefont {P.-A.}\ \bibnamefont
  {Malmqvist}}, \ and\ \bibinfo {author} {\bibfnamefont {B.~O.}\ \bibnamefont
  {Roos}},\ }\href@noop {} {\bibfield  {journal} {\bibinfo  {journal} {The
  Journal of chemical physics}\ }\textbf {\bibinfo {volume} {96}},\ \bibinfo
  {pages} {1218} (\bibinfo {year} {1992})}\BibitemShut {NoStop}%
\bibitem [{\citenamefont {Foresman}\ \emph {et~al.}(1992)\citenamefont
  {Foresman}, \citenamefont {Head-Gordon}, \citenamefont {Pople},\ and\
  \citenamefont {Frisch}}]{foresman1992toward}%
  \BibitemOpen
  \bibfield  {author} {\bibinfo {author} {\bibfnamefont {J.~B.}\ \bibnamefont
  {Foresman}}, \bibinfo {author} {\bibfnamefont {M.}~\bibnamefont
  {Head-Gordon}}, \bibinfo {author} {\bibfnamefont {J.~A.}\ \bibnamefont
  {Pople}}, \ and\ \bibinfo {author} {\bibfnamefont {M.~J.}\ \bibnamefont
  {Frisch}},\ }\href@noop {} {\bibfield  {journal} {\bibinfo  {journal} {The
  Journal of Physical Chemistry}\ }\textbf {\bibinfo {volume} {96}},\ \bibinfo
  {pages} {135} (\bibinfo {year} {1992})}\BibitemShut {NoStop}%
\bibitem [{\citenamefont {Stanton}\ and\ \citenamefont
  {Bartlett}(1993)}]{stanton1993equation}%
  \BibitemOpen
  \bibfield  {author} {\bibinfo {author} {\bibfnamefont {J.~F.}\ \bibnamefont
  {Stanton}}\ and\ \bibinfo {author} {\bibfnamefont {R.~J.}\ \bibnamefont
  {Bartlett}},\ }\href@noop {} {\bibfield  {journal} {\bibinfo  {journal} {The
  Journal of chemical physics}\ }\textbf {\bibinfo {volume} {98}},\ \bibinfo
  {pages} {7029} (\bibinfo {year} {1993})}\BibitemShut {NoStop}%
\bibitem [{\citenamefont {Angeli}\ \emph {et~al.}(2001)\citenamefont {Angeli},
  \citenamefont {Cimiraglia}, \citenamefont {Evangelisti}, \citenamefont
  {Leininger},\ and\ \citenamefont {Malrieu}}]{angeli2001introduction}%
  \BibitemOpen
  \bibfield  {author} {\bibinfo {author} {\bibfnamefont {C.}~\bibnamefont
  {Angeli}}, \bibinfo {author} {\bibfnamefont {R.}~\bibnamefont {Cimiraglia}},
  \bibinfo {author} {\bibfnamefont {S.}~\bibnamefont {Evangelisti}}, \bibinfo
  {author} {\bibfnamefont {T.}~\bibnamefont {Leininger}}, \ and\ \bibinfo
  {author} {\bibfnamefont {J.-P.}\ \bibnamefont {Malrieu}},\ }\href@noop {}
  {\bibfield  {journal} {\bibinfo  {journal} {The Journal of Chemical Physics}\
  }\textbf {\bibinfo {volume} {114}},\ \bibinfo {pages} {10252} (\bibinfo
  {year} {2001})}\BibitemShut {NoStop}%
\bibitem [{\citenamefont {Gilbert}\ \emph {et~al.}(2008)\citenamefont
  {Gilbert}, \citenamefont {Besley},\ and\ \citenamefont
  {Gill}}]{gilbert2008self}%
  \BibitemOpen
  \bibfield  {author} {\bibinfo {author} {\bibfnamefont {A.~T.}\ \bibnamefont
  {Gilbert}}, \bibinfo {author} {\bibfnamefont {N.~A.}\ \bibnamefont {Besley}},
  \ and\ \bibinfo {author} {\bibfnamefont {P.~M.}\ \bibnamefont {Gill}},\
  }\href@noop {} {\bibfield  {journal} {\bibinfo  {journal} {The Journal of
  Physical Chemistry A}\ }\textbf {\bibinfo {volume} {112}},\ \bibinfo {pages}
  {13164} (\bibinfo {year} {2008})}\BibitemShut {NoStop}%
\bibitem [{\citenamefont {Shea}\ and\ \citenamefont
  {Neuscamman}(2018)}]{shea2018communication}%
  \BibitemOpen
  \bibfield  {author} {\bibinfo {author} {\bibfnamefont {J.~A.}\ \bibnamefont
  {Shea}}\ and\ \bibinfo {author} {\bibfnamefont {E.}~\bibnamefont
  {Neuscamman}},\ }\href@noop {} {\bibfield  {journal} {\bibinfo  {journal}
  {The Journal of chemical physics}\ }\textbf {\bibinfo {volume} {149}}
  (\bibinfo {year} {2018})}\BibitemShut {NoStop}%
\bibitem [{\citenamefont {Tiago}\ and\ \citenamefont
  {Chelikowsky}(2006)}]{tiago2006optical}%
  \BibitemOpen
  \bibfield  {author} {\bibinfo {author} {\bibfnamefont {M.~L.}\ \bibnamefont
  {Tiago}}\ and\ \bibinfo {author} {\bibfnamefont {J.~R.}\ \bibnamefont
  {Chelikowsky}},\ }\href@noop {} {\bibfield  {journal} {\bibinfo  {journal}
  {Physical Review B—Condensed Matter and Materials Physics}\ }\textbf
  {\bibinfo {volume} {73}},\ \bibinfo {pages} {205334} (\bibinfo {year}
  {2006})}\BibitemShut {NoStop}%
\bibitem [{\citenamefont {Trofimov}\ and\ \citenamefont
  {Schirmer}(1995)}]{trofimov1995efficient}%
  \BibitemOpen
  \bibfield  {author} {\bibinfo {author} {\bibfnamefont {A.~B.}\ \bibnamefont
  {Trofimov}}\ and\ \bibinfo {author} {\bibfnamefont {J.}~\bibnamefont
  {Schirmer}},\ }\href@noop {} {\bibfield  {journal} {\bibinfo  {journal}
  {Journal of Physics B: Atomic, Molecular and Optical Physics}\ }\textbf
  {\bibinfo {volume} {28}},\ \bibinfo {pages} {2299} (\bibinfo {year}
  {1995})}\BibitemShut {NoStop}%
\bibitem [{\citenamefont {Runge}\ and\ \citenamefont
  {Gross}(1984)}]{runge1984density}%
  \BibitemOpen
  \bibfield  {author} {\bibinfo {author} {\bibfnamefont {E.}~\bibnamefont
  {Runge}}\ and\ \bibinfo {author} {\bibfnamefont {E.~K.}\ \bibnamefont
  {Gross}},\ }\href@noop {} {\bibfield  {journal} {\bibinfo  {journal}
  {Physical review letters}\ }\textbf {\bibinfo {volume} {52}},\ \bibinfo
  {pages} {997} (\bibinfo {year} {1984})}\BibitemShut {NoStop}%
\bibitem [{\citenamefont {Dreuw}\ and\ \citenamefont
  {Head-Gordon}(2005)}]{dreuw2005single}%
  \BibitemOpen
  \bibfield  {author} {\bibinfo {author} {\bibfnamefont {A.}~\bibnamefont
  {Dreuw}}\ and\ \bibinfo {author} {\bibfnamefont {M.}~\bibnamefont
  {Head-Gordon}},\ }\href@noop {} {\bibfield  {journal} {\bibinfo  {journal}
  {Chemical reviews}\ }\textbf {\bibinfo {volume} {105}},\ \bibinfo {pages}
  {4009} (\bibinfo {year} {2005})}\BibitemShut {NoStop}%
\bibitem [{\citenamefont {Hait}\ and\ \citenamefont
  {Head-Gordon}(2021)}]{hait2021orbital}%
  \BibitemOpen
  \bibfield  {author} {\bibinfo {author} {\bibfnamefont {D.}~\bibnamefont
  {Hait}}\ and\ \bibinfo {author} {\bibfnamefont {M.}~\bibnamefont
  {Head-Gordon}},\ }\href@noop {} {\bibfield  {journal} {\bibinfo  {journal}
  {The journal of physical chemistry letters}\ }\textbf {\bibinfo {volume}
  {12}},\ \bibinfo {pages} {4517} (\bibinfo {year} {2021})}\BibitemShut
  {NoStop}%
\bibitem [{\citenamefont {Goddard~III}\ \emph {et~al.}(1969)\citenamefont
  {Goddard~III} \emph {et~al.}}]{goddard1969excited}%
  \BibitemOpen
  \bibfield  {author} {\bibinfo {author} {\bibfnamefont {W.~A.}\ \bibnamefont
  {Goddard~III}} \emph {et~al.},\ }\href@noop {} {\bibfield  {journal}
  {\bibinfo  {journal} {Chemical Physics Letters}\ }\textbf {\bibinfo {volume}
  {3}},\ \bibinfo {pages} {414} (\bibinfo {year} {1969})}\BibitemShut {NoStop}%
\bibitem [{\citenamefont {Lee}\ \emph {et~al.}(2019)\citenamefont {Lee},
  \citenamefont {Small},\ and\ \citenamefont {Head-Gordon}}]{lee2019excited}%
  \BibitemOpen
  \bibfield  {author} {\bibinfo {author} {\bibfnamefont {J.}~\bibnamefont
  {Lee}}, \bibinfo {author} {\bibfnamefont {D.~W.}\ \bibnamefont {Small}}, \
  and\ \bibinfo {author} {\bibfnamefont {M.}~\bibnamefont {Head-Gordon}},\
  }\href@noop {} {\bibfield  {journal} {\bibinfo  {journal} {The Journal of
  chemical physics}\ }\textbf {\bibinfo {volume} {151}} (\bibinfo {year}
  {2019})}\BibitemShut {NoStop}%
\bibitem [{\citenamefont {Kossoski}\ and\ \citenamefont
  {Loos}(2023)}]{kossoski2023state}%
  \BibitemOpen
  \bibfield  {author} {\bibinfo {author} {\bibfnamefont {F.}~\bibnamefont
  {Kossoski}}\ and\ \bibinfo {author} {\bibfnamefont {P.-F.}\ \bibnamefont
  {Loos}},\ }\href@noop {} {\bibfield  {journal} {\bibinfo  {journal} {Journal
  of Chemical Theory and Computation}\ }\textbf {\bibinfo {volume} {19}},\
  \bibinfo {pages} {2258} (\bibinfo {year} {2023})}\BibitemShut {NoStop}%
\bibitem [{\citenamefont {Foulkes}\ \emph {et~al.}(2001)\citenamefont
  {Foulkes}, \citenamefont {Mitas}, \citenamefont {Needs},\ and\ \citenamefont
  {Rajagopal}}]{foulkes2001quantum}%
  \BibitemOpen
  \bibfield  {author} {\bibinfo {author} {\bibfnamefont {W.}~\bibnamefont
  {Foulkes}}, \bibinfo {author} {\bibfnamefont {L.}~\bibnamefont {Mitas}},
  \bibinfo {author} {\bibfnamefont {R.}~\bibnamefont {Needs}}, \ and\ \bibinfo
  {author} {\bibfnamefont {G.}~\bibnamefont {Rajagopal}},\ }\href@noop {}
  {\bibfield  {journal} {\bibinfo  {journal} {Reviews of Modern Physics}\
  }\textbf {\bibinfo {volume} {73}},\ \bibinfo {pages} {33} (\bibinfo {year}
  {2001})}\BibitemShut {NoStop}%
\bibitem [{\citenamefont {Schautz}\ \emph {et~al.}(2004)\citenamefont
  {Schautz}, \citenamefont {Buda},\ and\ \citenamefont
  {Filippi}}]{schautz2004excitations}%
  \BibitemOpen
  \bibfield  {author} {\bibinfo {author} {\bibfnamefont {F.}~\bibnamefont
  {Schautz}}, \bibinfo {author} {\bibfnamefont {F.}~\bibnamefont {Buda}}, \
  and\ \bibinfo {author} {\bibfnamefont {C.}~\bibnamefont {Filippi}},\
  }\href@noop {} {\bibfield  {journal} {\bibinfo  {journal} {The Journal of
  chemical physics}\ }\textbf {\bibinfo {volume} {121}},\ \bibinfo {pages}
  {5836} (\bibinfo {year} {2004})}\BibitemShut {NoStop}%
\bibitem [{\citenamefont {Drummond}\ \emph {et~al.}(2005)\citenamefont
  {Drummond}, \citenamefont {Williamson}, \citenamefont {Needs},\ and\
  \citenamefont {Galli}}]{drummond2005electron}%
  \BibitemOpen
  \bibfield  {author} {\bibinfo {author} {\bibfnamefont {N.}~\bibnamefont
  {Drummond}}, \bibinfo {author} {\bibfnamefont {A.}~\bibnamefont
  {Williamson}}, \bibinfo {author} {\bibfnamefont {R.}~\bibnamefont {Needs}}, \
  and\ \bibinfo {author} {\bibfnamefont {G.}~\bibnamefont {Galli}},\
  }\href@noop {} {\bibfield  {journal} {\bibinfo  {journal} {Physical review
  letters}\ }\textbf {\bibinfo {volume} {95}},\ \bibinfo {pages} {096801}
  (\bibinfo {year} {2005})}\BibitemShut {NoStop}%
\bibitem [{\citenamefont {Zimmerman}\ \emph {et~al.}(2009)\citenamefont
  {Zimmerman}, \citenamefont {Toulouse}, \citenamefont {Zhang}, \citenamefont
  {Musgrave},\ and\ \citenamefont {Umrigar}}]{zimmerman2009excited}%
  \BibitemOpen
  \bibfield  {author} {\bibinfo {author} {\bibfnamefont {P.~M.}\ \bibnamefont
  {Zimmerman}}, \bibinfo {author} {\bibfnamefont {J.}~\bibnamefont {Toulouse}},
  \bibinfo {author} {\bibfnamefont {Z.}~\bibnamefont {Zhang}}, \bibinfo
  {author} {\bibfnamefont {C.~B.}\ \bibnamefont {Musgrave}}, \ and\ \bibinfo
  {author} {\bibfnamefont {C.}~\bibnamefont {Umrigar}},\ }\href@noop {}
  {\bibfield  {journal} {\bibinfo  {journal} {The Journal of chemical physics}\
  }\textbf {\bibinfo {volume} {131}} (\bibinfo {year} {2009})}\BibitemShut
  {NoStop}%
\bibitem [{\citenamefont {Dubecky}\ \emph {et~al.}(2010)\citenamefont
  {Dubecky}, \citenamefont {Derian}, \citenamefont {Mitas},\ and\ \citenamefont
  {Stich}}]{dubecky2010ground}%
  \BibitemOpen
  \bibfield  {author} {\bibinfo {author} {\bibfnamefont {M.}~\bibnamefont
  {Dubecky}}, \bibinfo {author} {\bibfnamefont {R.}~\bibnamefont {Derian}},
  \bibinfo {author} {\bibfnamefont {L.}~\bibnamefont {Mitas}}, \ and\ \bibinfo
  {author} {\bibfnamefont {I.}~\bibnamefont {Stich}},\ }\href@noop {}
  {\bibfield  {journal} {\bibinfo  {journal} {The Journal of chemical physics}\
  }\textbf {\bibinfo {volume} {133}} (\bibinfo {year} {2010})}\BibitemShut
  {NoStop}%
\bibitem [{\citenamefont {Scemama}\ \emph {et~al.}(2018)\citenamefont
  {Scemama}, \citenamefont {Benali}, \citenamefont {Jacquemin}, \citenamefont
  {Caffarel},\ and\ \citenamefont {Loos}}]{scemama2018excitation}%
  \BibitemOpen
  \bibfield  {author} {\bibinfo {author} {\bibfnamefont {A.}~\bibnamefont
  {Scemama}}, \bibinfo {author} {\bibfnamefont {A.}~\bibnamefont {Benali}},
  \bibinfo {author} {\bibfnamefont {D.}~\bibnamefont {Jacquemin}}, \bibinfo
  {author} {\bibfnamefont {M.}~\bibnamefont {Caffarel}}, \ and\ \bibinfo
  {author} {\bibfnamefont {P.-F.}\ \bibnamefont {Loos}},\ }\href@noop {}
  {\bibfield  {journal} {\bibinfo  {journal} {The Journal of Chemical Physics}\
  }\textbf {\bibinfo {volume} {149}} (\bibinfo {year} {2018})}\BibitemShut
  {NoStop}%
\bibitem [{\citenamefont {Pineda~Flores}\ and\ \citenamefont
  {Neuscamman}(2019)}]{pineda2019excited}%
  \BibitemOpen
  \bibfield  {author} {\bibinfo {author} {\bibfnamefont {S.~D.}\ \bibnamefont
  {Pineda~Flores}}\ and\ \bibinfo {author} {\bibfnamefont {E.}~\bibnamefont
  {Neuscamman}},\ }\href@noop {} {\bibfield  {journal} {\bibinfo  {journal}
  {The Journal of Physical Chemistry A}\ }\textbf {\bibinfo {volume} {123}},\
  \bibinfo {pages} {1487} (\bibinfo {year} {2019})}\BibitemShut {NoStop}%
\bibitem [{\citenamefont {Filippi}\ \emph {et~al.}(2016)\citenamefont
  {Filippi}, \citenamefont {Assaraf},\ and\ \citenamefont
  {Moroni}}]{filippi2016simple}%
  \BibitemOpen
  \bibfield  {author} {\bibinfo {author} {\bibfnamefont {C.}~\bibnamefont
  {Filippi}}, \bibinfo {author} {\bibfnamefont {R.}~\bibnamefont {Assaraf}}, \
  and\ \bibinfo {author} {\bibfnamefont {S.}~\bibnamefont {Moroni}},\
  }\href@noop {} {\bibfield  {journal} {\bibinfo  {journal} {The Journal of
  chemical physics}\ }\textbf {\bibinfo {volume} {144}},\ \bibinfo {pages}
  {194105} (\bibinfo {year} {2016})}\BibitemShut {NoStop}%
\bibitem [{\citenamefont {Huron}\ \emph {et~al.}(1973)\citenamefont {Huron},
  \citenamefont {Malrieu},\ and\ \citenamefont {Rancurel}}]{Huron1973}%
  \BibitemOpen
  \bibfield  {author} {\bibinfo {author} {\bibfnamefont {B.}~\bibnamefont
  {Huron}}, \bibinfo {author} {\bibfnamefont {J.~P.}\ \bibnamefont {Malrieu}},
  \ and\ \bibinfo {author} {\bibfnamefont {P.}~\bibnamefont {Rancurel}},\
  }\href {\doibase 10.1063/1.1679199} {\bibfield  {journal} {\bibinfo
  {journal} {J. Chem. Phys.}\ }\textbf {\bibinfo {volume} {58}},\ \bibinfo
  {pages} {5745} (\bibinfo {year} {1973})},\ \Eprint
  {http://arxiv.org/abs/https://doi.org/10.1063/1.1679199}
  {https://doi.org/10.1063/1.1679199} \BibitemShut {NoStop}%
\bibitem [{\citenamefont {Holmes}\ \emph {et~al.}(2016)\citenamefont {Holmes},
  \citenamefont {Tubman},\ and\ \citenamefont {Umrigar}}]{Holmes2016b}%
  \BibitemOpen
  \bibfield  {author} {\bibinfo {author} {\bibfnamefont {A.~A.}\ \bibnamefont
  {Holmes}}, \bibinfo {author} {\bibfnamefont {N.~M.}\ \bibnamefont {Tubman}},
  \ and\ \bibinfo {author} {\bibfnamefont {C.~J.}\ \bibnamefont {Umrigar}},\
  }\href {\doibase 10.1021/acs.jctc.6b00407} {\bibfield  {journal} {\bibinfo
  {journal} {J. Chem. Theory Comput.}\ }\textbf {\bibinfo {volume} {12}},\
  \bibinfo {pages} {3674} (\bibinfo {year} {2016})},\ \bibinfo {note} {pMID:
  27428771},\ \Eprint
  {http://arxiv.org/abs/https://doi.org/10.1021/acs.jctc.6b00407}
  {https://doi.org/10.1021/acs.jctc.6b00407} \BibitemShut {NoStop}%
\bibitem [{\citenamefont {Tubman}\ \emph {et~al.}(2016)\citenamefont {Tubman},
  \citenamefont {Lee}, \citenamefont {Takeshita}, \citenamefont {Head-Gordon},\
  and\ \citenamefont {Whaley}}]{tubman2016deterministic}%
  \BibitemOpen
  \bibfield  {author} {\bibinfo {author} {\bibfnamefont {N.~M.}\ \bibnamefont
  {Tubman}}, \bibinfo {author} {\bibfnamefont {J.}~\bibnamefont {Lee}},
  \bibinfo {author} {\bibfnamefont {T.~Y.}\ \bibnamefont {Takeshita}}, \bibinfo
  {author} {\bibfnamefont {M.}~\bibnamefont {Head-Gordon}}, \ and\ \bibinfo
  {author} {\bibfnamefont {K.~B.}\ \bibnamefont {Whaley}},\ }\href@noop {}
  {\bibfield  {journal} {\bibinfo  {journal} {The Journal of chemical physics}\
  }\textbf {\bibinfo {volume} {145}},\ \bibinfo {pages} {044112} (\bibinfo
  {year} {2016})}\BibitemShut {NoStop}%
\bibitem [{\citenamefont {Robinson}\ \emph {et~al.}(2017)\citenamefont
  {Robinson}, \citenamefont {Pineda~Flores},\ and\ \citenamefont
  {Neuscamman}}]{robinson2017excitation}%
  \BibitemOpen
  \bibfield  {author} {\bibinfo {author} {\bibfnamefont {P.~J.}\ \bibnamefont
  {Robinson}}, \bibinfo {author} {\bibfnamefont {S.~D.}\ \bibnamefont
  {Pineda~Flores}}, \ and\ \bibinfo {author} {\bibfnamefont {E.}~\bibnamefont
  {Neuscamman}},\ }\href@noop {} {\bibfield  {journal} {\bibinfo  {journal}
  {The Journal of Chemical Physics}\ }\textbf {\bibinfo {volume} {147}}
  (\bibinfo {year} {2017})}\BibitemShut {NoStop}%
\bibitem [{\citenamefont {Dash}\ \emph {et~al.}(2019)\citenamefont {Dash},
  \citenamefont {Feldt}, \citenamefont {Moroni}, \citenamefont {Scemama},\ and\
  \citenamefont {Filippi}}]{dash2019excited}%
  \BibitemOpen
  \bibfield  {author} {\bibinfo {author} {\bibfnamefont {M.}~\bibnamefont
  {Dash}}, \bibinfo {author} {\bibfnamefont {J.}~\bibnamefont {Feldt}},
  \bibinfo {author} {\bibfnamefont {S.}~\bibnamefont {Moroni}}, \bibinfo
  {author} {\bibfnamefont {A.}~\bibnamefont {Scemama}}, \ and\ \bibinfo
  {author} {\bibfnamefont {C.}~\bibnamefont {Filippi}},\ }\href@noop {}
  {\bibfield  {journal} {\bibinfo  {journal} {Journal of chemical theory and
  computation}\ }\textbf {\bibinfo {volume} {15}},\ \bibinfo {pages} {4896}
  (\bibinfo {year} {2019})}\BibitemShut {NoStop}%
\bibitem [{\citenamefont {Zhang}\ and\ \citenamefont
  {Krakauer}(2003)}]{zhang2003quantum}%
  \BibitemOpen
  \bibfield  {author} {\bibinfo {author} {\bibfnamefont {S.}~\bibnamefont
  {Zhang}}\ and\ \bibinfo {author} {\bibfnamefont {H.}~\bibnamefont
  {Krakauer}},\ }\href@noop {} {\bibfield  {journal} {\bibinfo  {journal}
  {Physical review letters}\ }\textbf {\bibinfo {volume} {90}},\ \bibinfo
  {pages} {136401} (\bibinfo {year} {2003})}\BibitemShut {NoStop}%
\bibitem [{\citenamefont {Shee}\ \emph {et~al.}(2019)\citenamefont {Shee},
  \citenamefont {Rudshteyn}, \citenamefont {Arthur}, \citenamefont {Zhang},
  \citenamefont {Reichman},\ and\ \citenamefont
  {Friesner}}]{shee2019achieving}%
  \BibitemOpen
  \bibfield  {author} {\bibinfo {author} {\bibfnamefont {J.}~\bibnamefont
  {Shee}}, \bibinfo {author} {\bibfnamefont {B.}~\bibnamefont {Rudshteyn}},
  \bibinfo {author} {\bibfnamefont {E.~J.}\ \bibnamefont {Arthur}}, \bibinfo
  {author} {\bibfnamefont {S.}~\bibnamefont {Zhang}}, \bibinfo {author}
  {\bibfnamefont {D.~R.}\ \bibnamefont {Reichman}}, \ and\ \bibinfo {author}
  {\bibfnamefont {R.~A.}\ \bibnamefont {Friesner}},\ }\href@noop {} {\bibfield
  {journal} {\bibinfo  {journal} {Journal of chemical theory and computation}\
  }\textbf {\bibinfo {volume} {15}},\ \bibinfo {pages} {2346} (\bibinfo {year}
  {2019})}\BibitemShut {NoStop}%
\bibitem [{\citenamefont {Motta}\ and\ \citenamefont
  {Zhang}(2018)}]{motta2018ab}%
  \BibitemOpen
  \bibfield  {author} {\bibinfo {author} {\bibfnamefont {M.}~\bibnamefont
  {Motta}}\ and\ \bibinfo {author} {\bibfnamefont {S.}~\bibnamefont {Zhang}},\
  }\href@noop {} {\bibfield  {journal} {\bibinfo  {journal} {Wiley
  Interdisciplinary Reviews: Computational Molecular Science}\ }\textbf
  {\bibinfo {volume} {8}},\ \bibinfo {pages} {e1364} (\bibinfo {year}
  {2018})}\BibitemShut {NoStop}%
\bibitem [{\citenamefont {Shi}\ and\ \citenamefont
  {Zhang}(2021)}]{shi2021some}%
  \BibitemOpen
  \bibfield  {author} {\bibinfo {author} {\bibfnamefont {H.}~\bibnamefont
  {Shi}}\ and\ \bibinfo {author} {\bibfnamefont {S.}~\bibnamefont {Zhang}},\
  }\href@noop {} {\bibfield  {journal} {\bibinfo  {journal} {The Journal of
  Chemical Physics}\ }\textbf {\bibinfo {volume} {154}},\ \bibinfo {pages}
  {024107} (\bibinfo {year} {2021})}\BibitemShut {NoStop}%
\bibitem [{\citenamefont {Lee}\ \emph {et~al.}(2022)\citenamefont {Lee},
  \citenamefont {Pham},\ and\ \citenamefont {Reichman}}]{lee2022twenty}%
  \BibitemOpen
  \bibfield  {author} {\bibinfo {author} {\bibfnamefont {J.}~\bibnamefont
  {Lee}}, \bibinfo {author} {\bibfnamefont {H.~Q.}\ \bibnamefont {Pham}}, \
  and\ \bibinfo {author} {\bibfnamefont {D.~R.}\ \bibnamefont {Reichman}},\
  }\href@noop {} {\bibfield  {journal} {\bibinfo  {journal} {Journal of
  Chemical Theory and Computation}\ }\textbf {\bibinfo {volume} {18}},\
  \bibinfo {pages} {7024} (\bibinfo {year} {2022})}\BibitemShut {NoStop}%
\bibitem [{\citenamefont {Chang}\ \emph {et~al.}(2016)\citenamefont {Chang},
  \citenamefont {Rubenstein},\ and\ \citenamefont
  {Morales}}]{chang2016auxiliary}%
  \BibitemOpen
  \bibfield  {author} {\bibinfo {author} {\bibfnamefont {C.-C.}\ \bibnamefont
  {Chang}}, \bibinfo {author} {\bibfnamefont {B.~M.}\ \bibnamefont
  {Rubenstein}}, \ and\ \bibinfo {author} {\bibfnamefont {M.~A.}\ \bibnamefont
  {Morales}},\ }\href@noop {} {\bibfield  {journal} {\bibinfo  {journal}
  {Physical Review B}\ }\textbf {\bibinfo {volume} {94}},\ \bibinfo {pages}
  {235144} (\bibinfo {year} {2016})}\BibitemShut {NoStop}%
\bibitem [{\citenamefont {Lee}\ \emph {et~al.}(2020)\citenamefont {Lee},
  \citenamefont {Malone},\ and\ \citenamefont {Morales}}]{lee2020utilizing}%
  \BibitemOpen
  \bibfield  {author} {\bibinfo {author} {\bibfnamefont {J.}~\bibnamefont
  {Lee}}, \bibinfo {author} {\bibfnamefont {F.~D.}\ \bibnamefont {Malone}}, \
  and\ \bibinfo {author} {\bibfnamefont {M.~A.}\ \bibnamefont {Morales}},\
  }\href@noop {} {\bibfield  {journal} {\bibinfo  {journal} {Journal of
  chemical theory and computation}\ }\textbf {\bibinfo {volume} {16}},\
  \bibinfo {pages} {3019} (\bibinfo {year} {2020})}\BibitemShut {NoStop}%
\bibitem [{\citenamefont {Mahajan}\ \emph {et~al.}(2022)\citenamefont
  {Mahajan}, \citenamefont {Lee},\ and\ \citenamefont
  {Sharma}}]{mahajan2022selected}%
  \BibitemOpen
  \bibfield  {author} {\bibinfo {author} {\bibfnamefont {A.}~\bibnamefont
  {Mahajan}}, \bibinfo {author} {\bibfnamefont {J.}~\bibnamefont {Lee}}, \ and\
  \bibinfo {author} {\bibfnamefont {S.}~\bibnamefont {Sharma}},\ }\href@noop {}
  {\bibfield  {journal} {\bibinfo  {journal} {The Journal of Chemical Physics}\
  }\textbf {\bibinfo {volume} {156}} (\bibinfo {year} {2022})}\BibitemShut
  {NoStop}%
\bibitem [{\citenamefont {Mahajan}\ \emph {et~al.}(2025)\citenamefont
  {Mahajan}, \citenamefont {Thorpe}, \citenamefont {Kurian}, \citenamefont
  {Reichman}, \citenamefont {Matthews},\ and\ \citenamefont
  {Sharma}}]{mahajan2025beyond}%
  \BibitemOpen
  \bibfield  {author} {\bibinfo {author} {\bibfnamefont {A.}~\bibnamefont
  {Mahajan}}, \bibinfo {author} {\bibfnamefont {J.~H.}\ \bibnamefont {Thorpe}},
  \bibinfo {author} {\bibfnamefont {J.~S.}\ \bibnamefont {Kurian}}, \bibinfo
  {author} {\bibfnamefont {D.~R.}\ \bibnamefont {Reichman}}, \bibinfo {author}
  {\bibfnamefont {D.~A.}\ \bibnamefont {Matthews}}, \ and\ \bibinfo {author}
  {\bibfnamefont {S.}~\bibnamefont {Sharma}},\ }\href@noop {} {\bibfield
  {journal} {\bibinfo  {journal} {Journal of Chemical Theory and Computation}\
  }\textbf {\bibinfo {volume} {21}},\ \bibinfo {pages} {1626} (\bibinfo {year}
  {2025})}\BibitemShut {NoStop}%
\bibitem [{\citenamefont {Danilov}\ \emph {et~al.}(2025)\citenamefont
  {Danilov}, \citenamefont {Ganoe}, \citenamefont {Munyi},\ and\ \citenamefont
  {Shee}}]{danilov2025capturing}%
  \BibitemOpen
  \bibfield  {author} {\bibinfo {author} {\bibfnamefont {D.}~\bibnamefont
  {Danilov}}, \bibinfo {author} {\bibfnamefont {B.}~\bibnamefont {Ganoe}},
  \bibinfo {author} {\bibfnamefont {M.}~\bibnamefont {Munyi}}, \ and\ \bibinfo
  {author} {\bibfnamefont {J.}~\bibnamefont {Shee}},\ }\href@noop {} {\bibfield
   {journal} {\bibinfo  {journal} {Journal of Chemical Theory and Computation}\
  }\textbf {\bibinfo {volume} {21}},\ \bibinfo {pages} {1136} (\bibinfo {year}
  {2025})}\BibitemShut {NoStop}%
\bibitem [{\citenamefont {Xiao}\ \emph {et~al.}(2025)\citenamefont {Xiao},
  \citenamefont {Lu}, \citenamefont {Chen}, \citenamefont {Xiang},\ and\
  \citenamefont {Zhang}}]{xiao2025implementingadvancedtrialwave}%
  \BibitemOpen
  \bibfield  {author} {\bibinfo {author} {\bibfnamefont {Z.-Y.}\ \bibnamefont
  {Xiao}}, \bibinfo {author} {\bibfnamefont {Z.}~\bibnamefont {Lu}}, \bibinfo
  {author} {\bibfnamefont {Y.}~\bibnamefont {Chen}}, \bibinfo {author}
  {\bibfnamefont {T.}~\bibnamefont {Xiang}}, \ and\ \bibinfo {author}
  {\bibfnamefont {S.}~\bibnamefont {Zhang}},\ }\href
  {https://arxiv.org/abs/2505.18519} {\enquote {\bibinfo {title} {Implementing
  advanced trial wave functions in fermion quantum monte carlo via stochastic
  sampling},}\ } (\bibinfo {year} {2025}),\ \Eprint
  {http://arxiv.org/abs/2505.18519} {arXiv:2505.18519 [cond-mat.str-el]}
  \BibitemShut {NoStop}%
\bibitem [{\citenamefont {Purwanto}\ \emph {et~al.}(2009)\citenamefont
  {Purwanto}, \citenamefont {Zhang},\ and\ \citenamefont
  {Krakauer}}]{purwanto2009excited}%
  \BibitemOpen
  \bibfield  {author} {\bibinfo {author} {\bibfnamefont {W.}~\bibnamefont
  {Purwanto}}, \bibinfo {author} {\bibfnamefont {S.}~\bibnamefont {Zhang}}, \
  and\ \bibinfo {author} {\bibfnamefont {H.}~\bibnamefont {Krakauer}},\
  }\href@noop {} {\bibfield  {journal} {\bibinfo  {journal} {The Journal of
  chemical physics}\ }\textbf {\bibinfo {volume} {130}} (\bibinfo {year}
  {2009})}\BibitemShut {NoStop}%
\bibitem [{\citenamefont {Ma}\ \emph {et~al.}(2013)\citenamefont {Ma},
  \citenamefont {Zhang},\ and\ \citenamefont {Krakauer}}]{ma2013excited}%
  \BibitemOpen
  \bibfield  {author} {\bibinfo {author} {\bibfnamefont {F.}~\bibnamefont
  {Ma}}, \bibinfo {author} {\bibfnamefont {S.}~\bibnamefont {Zhang}}, \ and\
  \bibinfo {author} {\bibfnamefont {H.}~\bibnamefont {Krakauer}},\ }\href@noop
  {} {\bibfield  {journal} {\bibinfo  {journal} {New Journal of Physics}\
  }\textbf {\bibinfo {volume} {15}},\ \bibinfo {pages} {093017} (\bibinfo
  {year} {2013})}\BibitemShut {NoStop}%
\bibitem [{\citenamefont {Loos}\ \emph
  {et~al.}(2020{\natexlab{a}})\citenamefont {Loos}, \citenamefont {Scemama},\
  and\ \citenamefont {Jacquemin}}]{loos2020quest}%
  \BibitemOpen
  \bibfield  {author} {\bibinfo {author} {\bibfnamefont {P.-F.}\ \bibnamefont
  {Loos}}, \bibinfo {author} {\bibfnamefont {A.}~\bibnamefont {Scemama}}, \
  and\ \bibinfo {author} {\bibfnamefont {D.}~\bibnamefont {Jacquemin}},\
  }\href@noop {} {\bibfield  {journal} {\bibinfo  {journal} {The journal of
  physical chemistry letters}\ }\textbf {\bibinfo {volume} {11}},\ \bibinfo
  {pages} {2374} (\bibinfo {year} {2020}{\natexlab{a}})}\BibitemShut {NoStop}%
\bibitem [{\citenamefont {Motta}\ and\ \citenamefont
  {Zhang}(2017)}]{motta2017computation}%
  \BibitemOpen
  \bibfield  {author} {\bibinfo {author} {\bibfnamefont {M.}~\bibnamefont
  {Motta}}\ and\ \bibinfo {author} {\bibfnamefont {S.}~\bibnamefont {Zhang}},\
  }\href@noop {} {\bibfield  {journal} {\bibinfo  {journal} {Journal of
  chemical theory and computation}\ }\textbf {\bibinfo {volume} {13}},\
  \bibinfo {pages} {5367} (\bibinfo {year} {2017})}\BibitemShut {NoStop}%
\bibitem [{\citenamefont {Motta}\ \emph {et~al.}(2019)\citenamefont {Motta},
  \citenamefont {Zhang},\ and\ \citenamefont {Chan}}]{motta2019hamiltonian}%
  \BibitemOpen
  \bibfield  {author} {\bibinfo {author} {\bibfnamefont {M.}~\bibnamefont
  {Motta}}, \bibinfo {author} {\bibfnamefont {S.}~\bibnamefont {Zhang}}, \ and\
  \bibinfo {author} {\bibfnamefont {G.~K.-L.}\ \bibnamefont {Chan}},\
  }\href@noop {} {\bibfield  {journal} {\bibinfo  {journal} {Physical Review
  B}\ }\textbf {\bibinfo {volume} {100}},\ \bibinfo {pages} {045127} (\bibinfo
  {year} {2019})}\BibitemShut {NoStop}%
\bibitem [{\citenamefont {Mahajan}\ and\ \citenamefont
  {Sharma}(2020)}]{mahajan2020efficient}%
  \BibitemOpen
  \bibfield  {author} {\bibinfo {author} {\bibfnamefont {A.}~\bibnamefont
  {Mahajan}}\ and\ \bibinfo {author} {\bibfnamefont {S.}~\bibnamefont
  {Sharma}},\ }\href@noop {} {\bibfield  {journal} {\bibinfo  {journal} {The
  Journal of Chemical Physics}\ }\textbf {\bibinfo {volume} {153}},\ \bibinfo
  {pages} {194108} (\bibinfo {year} {2020})}\BibitemShut {NoStop}%
\bibitem [{\citenamefont {Mahajan}\ and\ \citenamefont
  {Sharma}(2021)}]{mahajan2021taming}%
  \BibitemOpen
  \bibfield  {author} {\bibinfo {author} {\bibfnamefont {A.}~\bibnamefont
  {Mahajan}}\ and\ \bibinfo {author} {\bibfnamefont {S.}~\bibnamefont
  {Sharma}},\ }\href@noop {} {\bibfield  {journal} {\bibinfo  {journal}
  {Journal of Chemical Theory and Computation}\ }\textbf {\bibinfo {volume}
  {17}},\ \bibinfo {pages} {4786} (\bibinfo {year} {2021})}\BibitemShut
  {NoStop}%
\bibitem [{tri()}]{triplet_eom}%
  \BibitemOpen
  \href@noop {} {}\bibinfo {note} {Note that CIS and EOM-CISD are not suitable
  triplet trial states in AFQMC because they have \(S_z=0\). Walkers with
  \(S_z=0\) generally collapse to singlets (\(S^2=0\)), causing vanishing
  overlaps with these triplet trials.}\BibitemShut {Stop}%
\bibitem [{\citenamefont {Jiang}\ \emph {et~al.}(2024)\citenamefont {Jiang},
  \citenamefont {O'Gorman}, \citenamefont {Mahajan},\ and\ \citenamefont
  {Lee}}]{jiang2024unbiasing}%
  \BibitemOpen
  \bibfield  {author} {\bibinfo {author} {\bibfnamefont {T.}~\bibnamefont
  {Jiang}}, \bibinfo {author} {\bibfnamefont {B.}~\bibnamefont {O'Gorman}},
  \bibinfo {author} {\bibfnamefont {A.}~\bibnamefont {Mahajan}}, \ and\
  \bibinfo {author} {\bibfnamefont {J.}~\bibnamefont {Lee}},\ }\href@noop {}
  {\bibfield  {journal} {\bibinfo  {journal} {arXiv preprint arXiv:2405.05440}\
  } (\bibinfo {year} {2024})}\BibitemShut {NoStop}%
\bibitem [{\citenamefont {Sun}\ \emph {et~al.}(2018)\citenamefont {Sun},
  \citenamefont {Berkelbach}, \citenamefont {Blunt}, \citenamefont {Booth},
  \citenamefont {Guo}, \citenamefont {Li}, \citenamefont {Liu}, \citenamefont
  {McClain}, \citenamefont {Sayfutyarova}, \citenamefont {Sharma},
  \citenamefont {Wouters},\ and\ \citenamefont {Chan}}]{sun2018pyscf}%
  \BibitemOpen
  \bibfield  {author} {\bibinfo {author} {\bibfnamefont {Q.}~\bibnamefont
  {Sun}}, \bibinfo {author} {\bibfnamefont {T.~C.}\ \bibnamefont {Berkelbach}},
  \bibinfo {author} {\bibfnamefont {N.~S.}\ \bibnamefont {Blunt}}, \bibinfo
  {author} {\bibfnamefont {G.~H.}\ \bibnamefont {Booth}}, \bibinfo {author}
  {\bibfnamefont {S.}~\bibnamefont {Guo}}, \bibinfo {author} {\bibfnamefont
  {Z.}~\bibnamefont {Li}}, \bibinfo {author} {\bibfnamefont {J.}~\bibnamefont
  {Liu}}, \bibinfo {author} {\bibfnamefont {J.~D.}\ \bibnamefont {McClain}},
  \bibinfo {author} {\bibfnamefont {E.~R.}\ \bibnamefont {Sayfutyarova}},
  \bibinfo {author} {\bibfnamefont {S.}~\bibnamefont {Sharma}}, \bibinfo
  {author} {\bibfnamefont {S.}~\bibnamefont {Wouters}}, \ and\ \bibinfo
  {author} {\bibfnamefont {K.-L.~G.}\ \bibnamefont {Chan}},\ }\href@noop {}
  {\bibfield  {journal} {\bibinfo  {journal} {WIREs Comput. Mol. Sci.}\
  }\textbf {\bibinfo {volume} {8}},\ \bibinfo {pages} {e1340} (\bibinfo {year}
  {2018})}\BibitemShut {NoStop}%
\bibitem [{dqm(2025)}]{dqmc_code}%
  \BibitemOpen
  \href@noop {} {}\bibinfo {howpublished}
  {\url{https://github.com/ankit76/ad_afqmc/}} (\bibinfo {year}
  {2025})\BibitemShut {NoStop}%
\bibitem [{\citenamefont {Sharma}\ \emph {et~al.}(2017)\citenamefont {Sharma},
  \citenamefont {Holmes}, \citenamefont {Jeanmairet}, \citenamefont {Alavi},\
  and\ \citenamefont {Umrigar}}]{ShaHolUmr}%
  \BibitemOpen
  \bibfield  {author} {\bibinfo {author} {\bibfnamefont {S.}~\bibnamefont
  {Sharma}}, \bibinfo {author} {\bibfnamefont {A.~A.}\ \bibnamefont {Holmes}},
  \bibinfo {author} {\bibfnamefont {G.}~\bibnamefont {Jeanmairet}}, \bibinfo
  {author} {\bibfnamefont {A.}~\bibnamefont {Alavi}}, \ and\ \bibinfo {author}
  {\bibfnamefont {C.~J.}\ \bibnamefont {Umrigar}},\ }\href {\doibase
  10.1021/acs.jctc.6b01028} {\bibfield  {journal} {\bibinfo  {journal} {J.
  Chem. Theory Comput.}\ }\textbf {\bibinfo {volume} {13}},\ \bibinfo {pages}
  {1595} (\bibinfo {year} {2017})},\ \bibinfo {note} {pMID: 28263594},\ \Eprint
  {http://arxiv.org/abs/https://doi.org/10.1021/acs.jctc.6b01028}
  {https://doi.org/10.1021/acs.jctc.6b01028} \BibitemShut {NoStop}%
\bibitem [{\citenamefont {Loos}\ \emph
  {et~al.}(2020{\natexlab{b}})\citenamefont {Loos}, \citenamefont {Lipparini},
  \citenamefont {Boggio-Pasqua}, \citenamefont {Scemama},\ and\ \citenamefont
  {Jacquemin}}]{loos2020mountaineering}%
  \BibitemOpen
  \bibfield  {author} {\bibinfo {author} {\bibfnamefont {P.-F.}\ \bibnamefont
  {Loos}}, \bibinfo {author} {\bibfnamefont {F.}~\bibnamefont {Lipparini}},
  \bibinfo {author} {\bibfnamefont {M.}~\bibnamefont {Boggio-Pasqua}}, \bibinfo
  {author} {\bibfnamefont {A.}~\bibnamefont {Scemama}}, \ and\ \bibinfo
  {author} {\bibfnamefont {D.}~\bibnamefont {Jacquemin}},\ }\href@noop {}
  {\bibfield  {journal} {\bibinfo  {journal} {Journal of Chemical Theory and
  Computation}\ }\textbf {\bibinfo {volume} {16}},\ \bibinfo {pages} {1711}
  (\bibinfo {year} {2020}{\natexlab{b}})}\BibitemShut {NoStop}%
\bibitem [{\citenamefont {Loos}\ \emph {et~al.}(2018)\citenamefont {Loos},
  \citenamefont {Scemama}, \citenamefont {Blondel}, \citenamefont {Garniron},
  \citenamefont {Caffarel},\ and\ \citenamefont
  {Jacquemin}}]{loos2018mountaineering}%
  \BibitemOpen
  \bibfield  {author} {\bibinfo {author} {\bibfnamefont {P.-F.}\ \bibnamefont
  {Loos}}, \bibinfo {author} {\bibfnamefont {A.}~\bibnamefont {Scemama}},
  \bibinfo {author} {\bibfnamefont {A.}~\bibnamefont {Blondel}}, \bibinfo
  {author} {\bibfnamefont {Y.}~\bibnamefont {Garniron}}, \bibinfo {author}
  {\bibfnamefont {M.}~\bibnamefont {Caffarel}}, \ and\ \bibinfo {author}
  {\bibfnamefont {D.}~\bibnamefont {Jacquemin}},\ }\href@noop {} {\bibfield
  {journal} {\bibinfo  {journal} {Journal of chemical theory and computation}\
  }\textbf {\bibinfo {volume} {14}},\ \bibinfo {pages} {4360} (\bibinfo {year}
  {2018})}\BibitemShut {NoStop}%
\bibitem [{\citenamefont {Schriber}\ \emph {et~al.}(2018)\citenamefont
  {Schriber}, \citenamefont {Hannon}, \citenamefont {Li},\ and\ \citenamefont
  {Evangelista}}]{schriber2018combined}%
  \BibitemOpen
  \bibfield  {author} {\bibinfo {author} {\bibfnamefont {J.~B.}\ \bibnamefont
  {Schriber}}, \bibinfo {author} {\bibfnamefont {K.~P.}\ \bibnamefont
  {Hannon}}, \bibinfo {author} {\bibfnamefont {C.}~\bibnamefont {Li}}, \ and\
  \bibinfo {author} {\bibfnamefont {F.~A.}\ \bibnamefont {Evangelista}},\
  }\href@noop {} {\bibfield  {journal} {\bibinfo  {journal} {Journal of
  chemical theory and computation}\ }\textbf {\bibinfo {volume} {14}},\
  \bibinfo {pages} {6295} (\bibinfo {year} {2018})}\BibitemShut {NoStop}%
\bibitem [{\citenamefont {Dallos}\ and\ \citenamefont
  {Lischka}(2004)}]{dallos2004systematic}%
  \BibitemOpen
  \bibfield  {author} {\bibinfo {author} {\bibfnamefont {M.}~\bibnamefont
  {Dallos}}\ and\ \bibinfo {author} {\bibfnamefont {H.}~\bibnamefont
  {Lischka}},\ }\href@noop {} {\bibfield  {journal} {\bibinfo  {journal}
  {Theoretical Chemistry Accounts}\ }\textbf {\bibinfo {volume} {112}},\
  \bibinfo {pages} {16} (\bibinfo {year} {2004})}\BibitemShut {NoStop}%
\bibitem [{\citenamefont {Daday}\ \emph {et~al.}(2012)\citenamefont {Daday},
  \citenamefont {Smart}, \citenamefont {Booth}, \citenamefont {Alavi},\ and\
  \citenamefont {Filippi}}]{daday2012full}%
  \BibitemOpen
  \bibfield  {author} {\bibinfo {author} {\bibfnamefont {C.}~\bibnamefont
  {Daday}}, \bibinfo {author} {\bibfnamefont {S.}~\bibnamefont {Smart}},
  \bibinfo {author} {\bibfnamefont {G.~H.}\ \bibnamefont {Booth}}, \bibinfo
  {author} {\bibfnamefont {A.}~\bibnamefont {Alavi}}, \ and\ \bibinfo {author}
  {\bibfnamefont {C.}~\bibnamefont {Filippi}},\ }\href@noop {} {\bibfield
  {journal} {\bibinfo  {journal} {Journal of chemical theory and computation}\
  }\textbf {\bibinfo {volume} {8}},\ \bibinfo {pages} {4441} (\bibinfo {year}
  {2012})}\BibitemShut {NoStop}%
\bibitem [{\citenamefont {Watson}\ and\ \citenamefont
  {Chan}(2012)}]{watson2012excited}%
  \BibitemOpen
  \bibfield  {author} {\bibinfo {author} {\bibfnamefont {M.~A.}\ \bibnamefont
  {Watson}}\ and\ \bibinfo {author} {\bibfnamefont {G.~K.-L.}\ \bibnamefont
  {Chan}},\ }\href@noop {} {\bibfield  {journal} {\bibinfo  {journal} {Journal
  of chemical theory and computation}\ }\textbf {\bibinfo {volume} {8}},\
  \bibinfo {pages} {4013} (\bibinfo {year} {2012})}\BibitemShut {NoStop}%
\bibitem [{\citenamefont {Dong}\ \emph {et~al.}(2019)\citenamefont {Dong},
  \citenamefont {Gagliardi},\ and\ \citenamefont {Truhlar}}]{dong2019nature}%
  \BibitemOpen
  \bibfield  {author} {\bibinfo {author} {\bibfnamefont {S.~S.}\ \bibnamefont
  {Dong}}, \bibinfo {author} {\bibfnamefont {L.}~\bibnamefont {Gagliardi}}, \
  and\ \bibinfo {author} {\bibfnamefont {D.~G.}\ \bibnamefont {Truhlar}},\
  }\href@noop {} {\bibfield  {journal} {\bibinfo  {journal} {Journal of
  Chemical Theory and Computation}\ }\textbf {\bibinfo {volume} {15}},\
  \bibinfo {pages} {4591} (\bibinfo {year} {2019})}\BibitemShut {NoStop}%
\bibitem [{\citenamefont {Tuckman}\ and\ \citenamefont
  {Neuscamman}(2023)}]{tuckman2023excited}%
  \BibitemOpen
  \bibfield  {author} {\bibinfo {author} {\bibfnamefont {H.}~\bibnamefont
  {Tuckman}}\ and\ \bibinfo {author} {\bibfnamefont {E.}~\bibnamefont
  {Neuscamman}},\ }\href@noop {} {\bibfield  {journal} {\bibinfo  {journal}
  {Journal of Chemical Theory and Computation}\ }\textbf {\bibinfo {volume}
  {19}},\ \bibinfo {pages} {6160} (\bibinfo {year} {2023})}\BibitemShut
  {NoStop}%
\bibitem [{\citenamefont {Liang}\ \emph {et~al.}(2022)\citenamefont {Liang},
  \citenamefont {Feng}, \citenamefont {Hait},\ and\ \citenamefont
  {Head-Gordon}}]{liang2022revisiting}%
  \BibitemOpen
  \bibfield  {author} {\bibinfo {author} {\bibfnamefont {J.}~\bibnamefont
  {Liang}}, \bibinfo {author} {\bibfnamefont {X.}~\bibnamefont {Feng}},
  \bibinfo {author} {\bibfnamefont {D.}~\bibnamefont {Hait}}, \ and\ \bibinfo
  {author} {\bibfnamefont {M.}~\bibnamefont {Head-Gordon}},\ }\href@noop {}
  {\bibfield  {journal} {\bibinfo  {journal} {Journal of chemical theory and
  computation}\ }\textbf {\bibinfo {volume} {18}},\ \bibinfo {pages} {3460}
  (\bibinfo {year} {2022})}\BibitemShut {NoStop}%
\bibitem [{\citenamefont {Bendikov}\ \emph {et~al.}(2004)\citenamefont
  {Bendikov}, \citenamefont {Duong}, \citenamefont {Starkey}, \citenamefont
  {Houk}, \citenamefont {Carter},\ and\ \citenamefont
  {Wudl}}]{bendikov2004oligoacenes}%
  \BibitemOpen
  \bibfield  {author} {\bibinfo {author} {\bibfnamefont {M.}~\bibnamefont
  {Bendikov}}, \bibinfo {author} {\bibfnamefont {H.~M.}\ \bibnamefont {Duong}},
  \bibinfo {author} {\bibfnamefont {K.}~\bibnamefont {Starkey}}, \bibinfo
  {author} {\bibfnamefont {K.}~\bibnamefont {Houk}}, \bibinfo {author}
  {\bibfnamefont {E.~A.}\ \bibnamefont {Carter}}, \ and\ \bibinfo {author}
  {\bibfnamefont {F.}~\bibnamefont {Wudl}},\ }\href@noop {} {\bibfield
  {journal} {\bibinfo  {journal} {Journal of the American Chemical Society}\
  }\textbf {\bibinfo {volume} {126}},\ \bibinfo {pages} {7416} (\bibinfo {year}
  {2004})}\BibitemShut {NoStop}%
\bibitem [{\citenamefont {Hachmann}\ \emph {et~al.}(2007)\citenamefont
  {Hachmann}, \citenamefont {Dorando}, \citenamefont {Avil{\'e}s},\ and\
  \citenamefont {Chan}}]{hachmann2007radical}%
  \BibitemOpen
  \bibfield  {author} {\bibinfo {author} {\bibfnamefont {J.}~\bibnamefont
  {Hachmann}}, \bibinfo {author} {\bibfnamefont {J.~J.}\ \bibnamefont
  {Dorando}}, \bibinfo {author} {\bibfnamefont {M.}~\bibnamefont {Avil{\'e}s}},
  \ and\ \bibinfo {author} {\bibfnamefont {G.~K.}\ \bibnamefont {Chan}},\
  }\href@noop {} {\bibfield  {journal} {\bibinfo  {journal} {The Journal of
  chemical physics}\ }\textbf {\bibinfo {volume} {127}} (\bibinfo {year}
  {2007})}\BibitemShut {NoStop}%
\bibitem [{\citenamefont {Hajgat{\'o}}\ \emph {et~al.}(2011)\citenamefont
  {Hajgat{\'o}}, \citenamefont {Huzak},\ and\ \citenamefont
  {Deleuze}}]{hajgato2011focal}%
  \BibitemOpen
  \bibfield  {author} {\bibinfo {author} {\bibfnamefont {B.}~\bibnamefont
  {Hajgat{\'o}}}, \bibinfo {author} {\bibfnamefont {M.}~\bibnamefont {Huzak}},
  \ and\ \bibinfo {author} {\bibfnamefont {M.~S.}\ \bibnamefont {Deleuze}},\
  }\href@noop {} {\bibfield  {journal} {\bibinfo  {journal} {The Journal of
  Physical Chemistry A}\ }\textbf {\bibinfo {volume} {115}},\ \bibinfo {pages}
  {9282} (\bibinfo {year} {2011})}\BibitemShut {NoStop}%
\bibitem [{\citenamefont {Siebrand}(1967)}]{siebrand1967radiationless}%
  \BibitemOpen
  \bibfield  {author} {\bibinfo {author} {\bibfnamefont {W.}~\bibnamefont
  {Siebrand}},\ }\href@noop {} {\bibfield  {journal} {\bibinfo  {journal} {The
  Journal of Chemical Physics}\ }\textbf {\bibinfo {volume} {47}},\ \bibinfo
  {pages} {2411} (\bibinfo {year} {1967})}\BibitemShut {NoStop}%
\bibitem [{\citenamefont {Burgos}\ \emph {et~al.}(1977)\citenamefont {Burgos},
  \citenamefont {Pope}, \citenamefont {Swenberg},\ and\ \citenamefont
  {Alfano}}]{burgos1977heterofission}%
  \BibitemOpen
  \bibfield  {author} {\bibinfo {author} {\bibfnamefont {J.}~\bibnamefont
  {Burgos}}, \bibinfo {author} {\bibfnamefont {M.}~\bibnamefont {Pope}},
  \bibinfo {author} {\bibfnamefont {C.~E.}\ \bibnamefont {Swenberg}}, \ and\
  \bibinfo {author} {\bibfnamefont {R.}~\bibnamefont {Alfano}},\ }\href@noop {}
  {\bibfield  {journal} {\bibinfo  {journal} {physica status solidi (b)}\
  }\textbf {\bibinfo {volume} {83}},\ \bibinfo {pages} {249} (\bibinfo {year}
  {1977})}\BibitemShut {NoStop}%
\bibitem [{\citenamefont {Angliker}\ \emph {et~al.}(1982)\citenamefont
  {Angliker}, \citenamefont {Rommel},\ and\ \citenamefont
  {Wirz}}]{angliker1982electronic}%
  \BibitemOpen
  \bibfield  {author} {\bibinfo {author} {\bibfnamefont {H.}~\bibnamefont
  {Angliker}}, \bibinfo {author} {\bibfnamefont {E.}~\bibnamefont {Rommel}}, \
  and\ \bibinfo {author} {\bibfnamefont {J.}~\bibnamefont {Wirz}},\ }\href@noop
  {} {\bibfield  {journal} {\bibinfo  {journal} {Chemical Physics Letters}\
  }\textbf {\bibinfo {volume} {87}},\ \bibinfo {pages} {208} (\bibinfo {year}
  {1982})}\BibitemShut {NoStop}%
\bibitem [{\citenamefont {Weber}\ \emph {et~al.}(2022)\citenamefont {Weber},
  \citenamefont {Vuong}, \citenamefont {Devlaminck}, \citenamefont {Shee},
  \citenamefont {Lee}, \citenamefont {Reichman},\ and\ \citenamefont
  {Friesner}}]{weber2022localized}%
  \BibitemOpen
  \bibfield  {author} {\bibinfo {author} {\bibfnamefont {J.~L.}\ \bibnamefont
  {Weber}}, \bibinfo {author} {\bibfnamefont {H.}~\bibnamefont {Vuong}},
  \bibinfo {author} {\bibfnamefont {P.~A.}\ \bibnamefont {Devlaminck}},
  \bibinfo {author} {\bibfnamefont {J.}~\bibnamefont {Shee}}, \bibinfo {author}
  {\bibfnamefont {J.}~\bibnamefont {Lee}}, \bibinfo {author} {\bibfnamefont
  {D.~R.}\ \bibnamefont {Reichman}}, \ and\ \bibinfo {author} {\bibfnamefont
  {R.~A.}\ \bibnamefont {Friesner}},\ }\href@noop {} {\bibfield  {journal}
  {\bibinfo  {journal} {Journal of Chemical Theory and Computation}\ }\textbf
  {\bibinfo {volume} {18}},\ \bibinfo {pages} {3447} (\bibinfo {year}
  {2022})}\BibitemShut {NoStop}%
\bibitem [{\citenamefont {Loos}\ \emph {et~al.}(2021)\citenamefont {Loos},
  \citenamefont {Comin}, \citenamefont {Blase},\ and\ \citenamefont
  {Jacquemin}}]{loos2021reference}%
  \BibitemOpen
  \bibfield  {author} {\bibinfo {author} {\bibfnamefont {P.-F.}\ \bibnamefont
  {Loos}}, \bibinfo {author} {\bibfnamefont {M.}~\bibnamefont {Comin}},
  \bibinfo {author} {\bibfnamefont {X.}~\bibnamefont {Blase}}, \ and\ \bibinfo
  {author} {\bibfnamefont {D.}~\bibnamefont {Jacquemin}},\ }\href@noop {}
  {\bibfield  {journal} {\bibinfo  {journal} {Journal of Chemical Theory and
  Computation}\ }\textbf {\bibinfo {volume} {17}},\ \bibinfo {pages} {3666}
  (\bibinfo {year} {2021})}\BibitemShut {NoStop}%
\end{thebibliography}
%

\end{document}